# Predicting the ages of galaxies with an artificial neural network

Laura. J. Hunt[1,2]★ Kevin. A. Pimbblet[1,2] and David. M. Benoit[1,2]

[1]*E.A. Milne Centre, Faculty of Science and Engineering, University of Hull, Cottingham Road, Kingston-upon-Hull HU6 7RX, UK*
[2]*Centre of Excellence for Data Science, AI, and Modelling (DAIM), University of Hull, Cottingham Road, Kingston-upon-Hull, HU6 7RX, UK*

**ABSTRACT**
We present a new method of predicting the ages of galaxies using a machine learning (ML) algorithm with the goal of providing an alternative to traditional methods. We aim to match the ability of traditional models to predict the ages of galaxies by training an artificial neural network (ANN) to recognize the relationships between the equivalent widths of spectral indices and the mass-weighted ages of galaxies estimated by the MAGPHYS model in data release 3 (DR3) of the Galaxy and Mass Assembly (GAMA) survey. We discuss the optimization of our hyperparameters extensively and investigate the application of a custom loss function to reduce the influence of errors in our input data. To quantify the quality of our predictions we calculate the mean squared error (MSE), mean absolute error (MAE) and $R^2$ score for which we find MSE = 0.020, MAE = 0.108 and $R^2$ = 0.530. We find our predicted ages have a similar distribution with standard deviation $\sigma_p = 0.182$ compared with the GAMA true ages $\sigma_t = 0.207$. This is achieved in approximately 23 s to train our ANN on an 11th Gen Intel Core i9-11900H running at 2.50 GHz using 32 GB of RAM. We report our results for when light-weighted ages are used to train the ANN, which improves the accuracy of the predictions. Finally, we detail an evaluation of our method relating to physical properties and compare with other ML techniques to encourage future applications of ML techniques in astronomy.

## 1 INTRODUCTION

The field of astronomy is inundated with vast amounts of data that is unable to be processed by humans alone. Automation is already implemented in many areas of the field; however, larger jumps in the efficiency of data processing must be made in order to accommodate the large amounts of observations and data being produced. One solution is to apply machine learning (ML) techniques to astrophysical problems. This involves training an ML algorithm to automatically recognize patterns within data sets in order to make predictions about unseen data (for a review of ML in astronomy please see: Baron 2019; Smith & Geach 2023). While this may act as an alternative method to circumvent the laborious process of modelling and analysis by providing tools to process data and analyse the patterns within observations to draw new conclusions, it is important to note that traditional methods are still a vital part of the process of analysis. ML algorithms may offer additional speed for processing but traditional models have a longer history and therefore the science and systematics behind them are better understood. However, for the purpose of quickly processing a data set, ML algorithms should be utilized further such that we are able to characterize them in a similar manner to traditional models and understand the science they are built upon.

Artificial neural networks (ANNs) are a ML technique used most commonly for supervised classification. They were first implemented in 1992 to study galaxies by Storrie-Lombardi et al. (1992) as a method to predict morphological classification based on the physical properties of galaxies. This process entails visual classification of a training set of galaxies provided by researchers alongside calculation of 13 observable parameters such as surface brightness and measures of asymmetry. The ANN correctly classifies 64 per cent of galaxies, in comparison ESO AUTO, a non-ANN automated classification method, only correctly classifies 56 per cent. In subsequent years the classification of galaxies with ANNs has been improved and reviewed extensively with comparison to human based visual classification. Some methods of classification include classifying Hubble types with an ANN (Adams & Woolley 1994; Naim 1994; Lahav et al. 1995; Lahav 1997; Odewahn 1997; Goderya & Lolling 2002; Ball et al. 2004) and using Galaxy Zoo data (Banerji et al. 2010). Other classifications include stellar spectral classification using ANNs (Gulati et al. 1994), in the UV (Gulati et al. 1996) and for low-signal-to-noise (Folkes, Lahav & Maddox 1996).

The application of ANNs to astrophysical problems is not limited to classification as one of the major benefits of ANNs is their flexibility. In the study of galaxies, ANNs and multilayer perceptrons (MLPs) are able to predict properties such as star formation rate (SFR; Ellison et al. 2016) and photometric redshifts based on spectral energy distributions (SEDs) (Firth, Lahav & Somerville 2003; Vanzella et al. 2004; Brescia et al. 2014; Bilicki et al. 2018). However, in the field of galactic astronomy there is a noticeable lack of galaxy age estimations with ANNs.

A galaxy's integrated light spectra provides insight into its underlying processes and properties. Historically, the study of spectral evolution stems from the analysis of full integrated light spectra; however, recent studies of the emission and absorption of various spectral lines allow us to infer these properties with less free parameters. Some of these methods involve using models based on the stellar population synthesis (SPS) technique which is based on the idea that the stellar components of galaxies evolve on evolutionary tracks called isochrones that dictate the way in which their spectra evolves (e.g. Tinsley 1968; Spinrad & Taylor 1971; Faber 1972; Tinsley 1972; O'Connell 1976; Tinsley & Gunn 1976; Bruzual A. 1983; Pickles 1985; Rose 1985; Bruzual & Charlot 2003). More recent methods include the flexible stellar population synthesis (FSPS) model that integrates essential aspects of SPS in a flexible manner so different sets of isochrones and stellar spectral libraries may be used (Conroy, Gunn & White 2009; Conroy & Gunn 2010). MILES (Medium resolution INT Library of Empirical Spectra) models are able to extend from intermediate ages to much older ages by incorporating empirical properties and extensive photometric libraries rather than just stellar spectra (Vazdekis et al. 2010). Extended-MILES (E-MILES) models (Vazdekis et al. 2016) are UV extended SPS models that provide better resolution, stellar parameter coverage, and signal-to-noise ratio (SNR) by using the next-generation spectral library (Gregg et al. 2006). MaStar (MaNGA stellar library (Yan et al. 2019) stellar population models (Maraston et al. 2020) are capable of predicted SEDs for stellar populations of various chemical compositions and ages. These approaches take the stellar initial mass function (IMF), SFR and sometimes the chemical enrichment abundance to determine the integrated spectral evolution of the stellar population. Important properties of galaxies such as age, metallicity, and abundance ratios affect the line-strength indices present in their spectra (Faber 1973; Worthey 1994; Bressan, Chiosi & Tantalo 1996; Vazdekis et al. 1996; Cardiel et al. 1998). Equivalent widths (EWs) are commonly used to quantify spectral lines as they measure the fraction of energy removed from the spectrum by the line rather the height or position of the line (Spitzer 1978), for which we use the definition of EWs described by Cardiel et al. (1998)

★ E-mail: hunt.ljh@gmail.com

$$W_\lambda(\text{Å}) = \int_{time} (1 - S(\lambda)/C(\lambda)) \, d\lambda, \quad (1)$$

where $S(\lambda)$ is the observed spectrum and $C(\lambda)$ is the local continuum usually found through interpolation of $S(\lambda)$ between two adjacent spectral regions. Specific EWs are tracers for specific processes within the stellar population such as starbursts and ongoing star formation (e.g. Worthey & Ottaviani 1997; Bruzual & Charlot 2003; Sánchez Almeida et al. 2012; Moresco et al. 2018). For this reason, it is possible to estimate the ages of galaxies based on the their spectral information. Galactic ages can be defined as the median mass-weighted age of the stellar population. Mass-weighted age is defined by Citro et al. (2016) based on the definition of Gallazzi et al. (2005)

$$\langle t \rangle_{mass} = \frac{\int_0^t SFR(t-t')M(t')t' \, dt'}{\int_0^t SFR(t-t')M(t') \, dt'}, \quad (2)$$

where $SFR(t-t')$ is the SFR at time $(t-t')$ when the star was formed, $M(t')$ is the stellar mass given by a single-stellar population (SSP) of age $t'$.

A number of EWs are commonly used as tracers for star formation such as $D_n4000$ (Hamilton 1985), H$\alpha$, H$\beta$ (Worthey et al. 1994), H$\gamma_A$ (Worthey & Ottaviani 1997), but other indices are also associated with other processes that may indirectly be associated with stellar age. Line indices can be used as metallicity indicators which in turn may be related back to age, with the caveat that there are many factors that relate to both. It has been long since established that there are particular strong features in the spectra of late-type galaxies such as CH G band (G) features, the magnesium b triplet (MgG) (e.g. Vazdekis et al. 1996; Jørgensen 1999), the magnesium hydride trough (MH), the sodium D doublet (NaD) (Faber 1973; Brodie & Hanes 1986; Brodie & Huchra 1990). Whereas other indices such as H$\alpha$, H$\beta$, [O III] and [S II] are associated with the star formation, the presence of AGN, Seyferts, and LINERs (Baldwin, Phillips & Terlevich 1981; Kauffmann et al. 2003b; Kewley et al. 2006; Cid Fernandes et al. 2010).

There are already various models that use integrated light spectra or SEDs to determine the physical properties of galaxies. Optimization of these models to run quickly through large amounts of data is paramount to the future of galaxy evolution. MOPED is a Multiple Optimized Parameter Estimation and Data compression algorithm described by Heavens, Jimenez & Lahav (2000) that is able to recover physical parameters from galaxy spectra such as emission and absorption lines for which physical properties such as SFR may be determined (Reichardt, Jimenez & Heavens 2001). They describe a method of linear compression for data sets that are dependent on multiple parameters. This method is aimed at galaxy spectra as they are based on a few parameters such as age and SFR, etc. MOPED is able to take an entire spectra and compress it into this handful of useful parameters. Before MOPED, methods involved the estimation of single parameters; whereas, MOPED not only enables multiple parameter estimations but reduces the error of previous compression systems such Principal Component Analysis (PCA) whilst also being faster to compute.

STARLIGHT (Cid Fernandes et al. 2005) is a spectral synthesis model that is able to recover information such as stellar ages and stellar metallicities from observed galaxy spectra. They achieve this by fitting spectra with a linear combination of simple theoretical stellar populations computed with evolutionary synthesis models at the same spectral resolution as that of the SDSS. This involves a mix of computational techniques developed for empirical population synthesis but applied with aspects of evolutionary synthesis models. They use STARLIGHT to produce a catalogue of properties for 50 000 SDSS DR2 galaxies with an increased computing speed when compared with MOPED.

Ocvirk et al. (2006) describes a method called STEllar Content via Maximum A Posteriori (STECMAP) which is based on a non-parametric inversion for analysis of integrated light spectra based on the synthetic spectra of SSPs. Their main aim is to recover star formation history (SFH) and stellar age–metallicity relationships for galaxies. STECMAP has a non-parametric approach in order to avoid constraints on the shape of the distribution for derived properties such as stellar age distribution. With this method they find that STECMAP is not easily able to recover age estimations for the optical range of spectra no matter the spectral resolution.

Tojeiro et al. (2007) describes a direct improvement from MOPED that involves a method of VErsatile SPectral Analysis (VESPA) that is able to recover properties such as SFH and metallicity histories from galactic spectra. VESPA differs from previous models as it can adapt the number of parameters recovered from a given spectrum depending on its SNR, wavelength coverage, and presence of a young stellar population whilst again improving computational time. Tojeiro et al. (2007) estimates VESPA reduces computational time from 170 yr for MOPED to process the entire SDSS DR5 to just 1 yr for VESPA.

A brief summary of various more recent, commonly used SED fitting methods includes the method of Bayesian SED fitting called P12 developed by Pacifici et al. (2012) that can take into account the combination of stellar and nebulae emission from galaxies across a broad range of wavelengths. SpeedyMC (Acquaviva, Gawiser &

Guaita 2012) is based on the SED fitting software GalMC which is a Markov Chain Monte Carlo algorithm (Acquaviva et al. 2011). SpeedyMC uses pre-computed template libraries which makes it possible to run very quickly, even on a laptop.

BayEsian Analysis of gaLaxy sEds (or BEAGLE) also uses a combination of the MULTINEST algorithm and a flexible, fully self-consistent physical model in the UV to the NIR to model any combination of photometric and spectroscopic observables including galaxy age (Chevallard & Charlot 2016). AGNFitter fits SEDs of AGN between submillimeter and UV using a fully Bayesian Markov Chain Monte Carlo method (Calistro Rivera et al. 2016).

The Dense Basis method of SED fitting is also based in PYTHON minimizes bias and reduces scatter caused by SFH parametrization by using four different functional families to create a basis of SFHs from their combinations in order to determine an optimal number of SFH components statistically (Iyer & Gawiser 2017; Iyer et al. 2019). Another PYTHON based method of SED fitting is Prospector (Leja et al. 2017; Johnson et al. 2021). They use a flexible method to derive stellar population parameters from photometry and spectroscopy across UV to IR.

Bayesian Analysis of Galaxies for Physical Inference and Parameter EStimation (BAGPIPES) is a PYTHON tool that can generate complex model spectra for galaxies using spectroscopic and photometric data (Carnall et al. 2018, 2019a). BAGPIPES uses Bayesian fitting to model emission from FUV to microwave regimes then fits these models with the MULTINEST nested sampling algorithm (Feroz & Hobson 2008; Feroz, Hobson & Bridges 2009; Feroz et al. 2019) to varying spectroscopic and photometric observations. Code Investigating GALaxy (CIGALE) is another PYTHON code that uses a Bayesian based method that incorporates FUV to radio to derive physical properties from SEDs of galaxies (Boquien et al. 2019).

MIRKWOOD uses an ensemble of supervised ML models to bypass computationally heavy Bayesian-based SED fitting (Gilda, Lower & Narayanan 2021). It is trained on mock SEDs generated by galaxy formation simulations in which the physical properties are known which allows the MIRKWOOD algorithms to derive an accurate relationship between inputted photometry and the physical properties of galaxies. The PRObabilistic ValueAdded BGS (PROVABGS) Bayesian SED modelling framework used on the DESI Bright Galaxy Survey (BGS; (Hahn et al. 2023; Myers et al. 2023)) photometry and spectroscopy can be used to derive physical properties such as mass-weighted age. PROVABGS uses non-parametric SFH and metallicity history prescriptions to model SEDs with SPS. For a more in-depth review and comparison of recent SED fitting methods please see Pacifici et al. (2023).

Considering such massive data sets as SDSS, it is imperative that new, faster methods of analysis are developed for galaxy spectra. Ucci et al. (2017, 2018) describes a supervised ML algorithm that calculates physical properties of galaxies based on their emission-line spectra with a combination of AdaBoost and Decision Trees, called GAME (GAlaxy Machine learning for Emission lines). They are able to train the algorithm in approximately 10 min on a set size of $3 \times 10^4$ spectra. Once trained the model is able to predict the density, metallicity, column density, and ionization parameters of a single spectra in less than a few seconds, to make predictions for the entirety of the SDSS DR5 it would take approximately 417 h. Compared with traditional models this is a vast improvement in processing time.

Liew-Cain et al. (2021) describes a method to recover the age and metallicities of galaxies from SEDs using a convolutional neural network (CNN). CNNs use convolutional filters that are able to pass over data arrays, reducing their size in a specific way in order to derive patterns that the network is then able to decide is important or not important with backpropagation. This process allows the network to relate patterns in the input data to a given output, in this case age or metallicity. Once the network is trained it is then able to use these patterns to predict the outcomes of new input data. This paper is a successful proof of concept for which CNNs are found to robustly predict age and metallicity from the SEDs of galaxies.

The process of analysis can be further sped up with the use of photometric data as this does not require expensive and time-consuming spectroscopic observations and generally has better SNRs and less calibration systematic errors. However, this is at the expense of losing more subtle signals that allow us to break the degeneracy between metallicity and dust to constrain age. Li et al. (2022) uses a CNN with the similar goal of estimating physical properties of galaxies; however, they achieve this with photometric data rather than SEDs. They train their algorithm, called Painting IntrinsiC Attributes onto SDSS Objects (PICASSO), with multiband photometric images to reconstruct 2D maps of stellar mass, metallicity, age and gas mass, gas metallicity, and SFR.

A simpler ML algorithm such as an ANN could reduce this overall time to make predictions even further if used to predict less parameters. For example, as CNNs generally use 2D input data such as images, spectra or SEDs, the number of input features can be significantly larger than a simple ANN as even a small 50×50 image would require 2500 input features. Though a slightly longer training time would not make a significant impact on the overall computational time, the time it would take to predict outputs for a large data set of 2D images after training is completed could significantly increase this time. Though 1D data like galaxy spectra may be used in CNNs like those described by Lovell et al. (2019) that predicts star formation histories based on synthetic galaxy spectra generated by two cosmological hydrodynamic simulations, EAGLE (Schaye et al. 2015) and Illustris (Genel et al. 2014). Therefore, we provide the proof of concept of an ANN[1] that is able to predict the ages of galaxies based on the EWs of their spectra with the aim matching the predictive power of more traditional models but with a shorter computing time. We aim to encourage future studies that incorporate faster and simpler ML techniques to predict smaller numbers of properties that would otherwise take many hours worth of simulations to predict. The use of more simple ML algorithms should be more widely utilized by the field as a whole as an additional tool for data processing and analysis rather than only researchers with specialisms in ML. Therefore, we aim to show that a simple ANN coded with `Tensorflow Keras` could be used by any researchers to predict the ages of galaxies based on their EWs. Our paper is structured as follows: in Section 2 we describe the data set, this includes a description of data cleaning, feature selection and drawbacks of our data set. In Section 3 we describe the framework of the ANN we employ, this includes hyperparameter tuning. We report our results in Section 4 in which we describe different techniques aimed at improving the overall performance of the predictions. We discuss the results further in Section 5 and provide comparison other ML methods and the use of light-weighted ages instead of mass-weighted ages.

We assume $H_0 = 68 \text{ km s}^{-1} \text{ Mpc}^{-1}$, $\Omega_M = 0.31$, and $\Omega_\Lambda = 0.69$, in concordance with $\Lambda$CDM (Planck Collaboration et al. 2020).

---

[1] E. A. Milne Centre for Astrophysics GitHub: https://github.com/Milne-Centre

## 2 INPUT DATA

Our data is sourced from the GAlaxy and Mass Assembly (GAMA) survey (Driver et al. 2009; Liske et al. 2015). GAMA is a spectroscopic and photometric survey that spans ~300 000 galaxies over ~286 deg$^2$ up to r < 19.8 mag. From data release 3 (DR3) (Baldry et al. 2018), we use median mass-weighted ages generated by SED fitting programme MAGPHYS v06, which is described in full by da Cunha, Charlot & Elbaz (2008), as opposed to synthetic observations or ages derived from other models because MAGPHYS is a physically motivated model that consistently is able to interpret galaxy emission across ultraviolet, optical, and infrared wavelengths. These are median ages as the MAGPHYS model produces percentile estimations for mass-weighted ages such that each observation has 16–84th and 2.5–97.5th percentile ranges with a median age with these ranges. To reiterate, our true ages have an associated error as they are estimations within a given range. Therefore, the aim of our ANN is predict these 'true' ages as accurately as possible in an attempt to match the age estimations from GAMA however there will always be a limit on the ANNs performance as it could be making more accurate predictions to the real true ages but not the GAMA true age estimations. It is important to note that MAGPHYS use exponentially declined SFH models which have known issues in predicting ages (e.g. Carnall et al. 2019b; Lower et al. 2020) and as such the bias for the true ages must be estimated to be approximately >0.2 dex as the distribution of true values is affected by this. MAGPHYS follows the methods of Kauffmann et al. (2003a) to parametrize the star formation histories from a stellar library by characterizing an underlying continuous model by an age $t_g$ and a star formation time-scale parameter $\gamma$ and introduce random bursts on the continuous model. They use models with exponentially declining SFRs

$$\psi(t) \propto \exp(-\gamma t), \quad (3)$$

where $\gamma$ is the star formation time-scale parameter which corresponds to models with $\gamma$ = 0, 0.07 and 0.25 Gyr$^{-1}$ at ages $t$ = 1.4, 10, and 10 Gyr which represent starburst, normal star-forming, and quiescent star-forming galaxies. Biases may be introduced by their attempt to avoid oversampling galaxies with negligible current star formation and the inclusion of random bursts that occur with equal probability at any given time until $t_g$. In addition, they state that the likelihood of a galaxy having experienced a burst in the last 2 Gyr is set to 50 per cent.

In addition, we use EW measurements of absorption and emission lines and from the DirectSummation table in the SpecLineSFR v05 DMU (Gordon et al. 2017). A number of steps are taken to build our sample. First, we use galaxies that are observed with GAMA using survey code 5. We then discard observations that do not have corresponding SPECIDs in both the MAGPHYS v06 and SpecLineSFR v05 DMUs. We follow the recommendation of GAMA to use a number of given parameters in the SpecLineSFR v05 to ensure each observation has been reliably detected. These recommendations include using NQ≥4 and SN≥3 which yields a sample of 54 473 galaxies.

We take further steps to clean our data by calculating SNRs for each observation for each EW. We use the EW errors provided in the DirectSummation table of SpecLineSFR v05 to calculate SNRs for every EW for every galaxy. We then rank each EW based on how many observations have a good SNR (SNR ≥3) to provide a reliable sample. 24 of the original 51 EWs are chosen based on their SNR, however, we run the network a number of times to determine the best EWs for training the network. We determine this by removing each EW in turn and training the network with the remaining 23. We then confirm the performance of the EWs by starting with one EW then adding the better performing EWs one by one, retraining the network each time until the performance starts to decrease. We take these steps to find the best EWs for our data set to take into account the different SNRs for each observation as some EWs that are more associated with age. We find the best performance with the 14 chosen EWs that are illustrated in Table 1. For definitions of each EW please refer to (Gordon et al. 2017).

**Table 1.** To reduce errors when predicting age we use EWs that have a high proportion of observations with good SNRs. Out of a total 51 EWs provided by GAMA we choose the top 24. We then narrow this down to 14 that have the best predictive performance.

| EW | Count | EW (cont.) | Count (cont.) |
|---|---|---|---|
| $D_n 4000$ | 54 473 | NaD | 18 831 |
| H$\alpha$ | 37 238 | [OIII] R | 15 262 |
| MH | 30 239 | FC | 14 353 |
| [S II] B | 29 019 | [OIII] B | 12 049 |
| [S II] R | 25 199 | CNB | 11 542 |
| G | 22 304 | H$\gamma_A$ | 11 452 |
| MgG | 19 043 | H$\beta$ | 7957 |

Finally, with this sample we limit the age distribution to maintain uniformity across our sample. We do this to prevent the ANN from being trained to predict the most common age. If the distribution of ages is even across the sample then the ANN will be forced to train on the patterns across EWs rather than predicting ages based on the most common and therefore the most probable age. We want the network to find the most probable age based on the EWs for each galaxy, not the most statistically probable age based on the distribution of input ages. This is seen more commonly in classification problems that have very unbalanced data sets that cause the network to predict whichever class there happens to be more of without considering the data itself. If a network is only trained on 10 000 galaxies of similar ages and 100 with much younger/older ages, it will predict the most probable outcome based on the distribution of ages in the training data and predict an age similar to those 10 000 galaxies because it is 99 per cent likely to be correct. However, this does not mean the network has learned the relationship between the EWs of galaxies and their associated ages. Generally, for regression problems we see that the network will predict the mean of the training set if it is not balanced. Balancing the training set also means we can verify whether the network has correctly found the most probable age by comparing the means of the unbalanced validation set and its corresponding predicted ages.

## 3 ANN

An ANN is a supervised ML algorithm that consists of a fully connected set of layers including an input layer, hidden layers and an output layer. A supervised ML algorithm is given a set of training samples and accompanying labels in order to predict the relationship between a sample and its given label in order to make future predictions. ANNs are simple to create and train using PYTHON package Tensorflow Keras, for this reason they are an ideal candidate for smaller studies in which the relationship between different parameters needs to be explored. Our aim for this work is to determine whether a ML algorithm is able to find the relationship between EWs and galaxy ages, therefore a simple ANN provides proof of concept for further work into this relationship. ANN architecture is based on the neurons in human brains (McCulloch & Pitts 1943; Hopfield & Tank 1986). The network is trained by learning the patterns between a set of input features ($x_1, x_2,..., x_i,$) and their respective output $y$. It learns this pattern by updating the weights and biases of the nodes through a process called backpropagation in which the nodes and weights are tuned to reduce the output error as much as possible in order to produce the most accurate predictions. There are a number of factors that go into this process that optimize the performance of the network such as increasing the number of hidden layers, changing the number of nodes in each layer and tuning the hyperparameters. In order to evaluate the performance of the different ANN architecture and hyperparameters we use a set of evaluation metrics to quantify the quality of the network. For each hyperparameter we tune, we run the network 10 times with 40 epochs during each training phase in order to calculate averages values for our evaluation metrics.

### 3.1 Evaluation metrics

The evaluation metrics we use are chosen based on the data we are working with. As our data is fully numerical and we are using a regression ANN we choose mean squared error (MSE), mean absolute error (MAE) and the coefficient of determination score also known as the R-squared score ($R^2$).

MSE is a metric used to compare actual values against predicted values by computing the mean of the squares of the errors between the values (Bickel & Doksum 2015). For our ANN we use the MSE metric included with the `Tensorflow Keras` package which describes MSE as

$$MSE = \frac{1}{n} \sum_{i=1}^{n} (y_i - \hat{y}_i)^2, \quad (4)$$

where $n$ is equal to the total number of data points, $y_i$ is the actual value, and $\hat{y}_i$ is the predicted value. The closer to zero the calculated MSE is, the more accurate the predictions are.

MAE is similar to MSE as it also measures the errors between actual and predicted values, as shown

$$MAE = \frac{1}{n} \sum_{i=1}^{n} |y_i - \hat{y}_i|, \quad (5)$$

where $\hat{y}_i$ is the predicted value, $y_i$ is the corresponding true value, and $n$ is the number of data points. Similarly to MSE, the closer to zero the more accurate the predictions are. We choose to use both of these metrics as even though they are similar they show slight nuances about the prediction ability of the ANN. The MAE value is less sensitive to large errors in prediction whereas MSE is able to penalize this more.

The $R^2$ score, also known as the coefficient of determination, quantifies the ability of a model to predict the y values of an unseen data set through the proportion of explained variance

$$R^2 = 1 - \frac{\sum_{i=1}^{n}(y_i - \hat{y}_i)^2}{\sum_{i=1}^{n}(y_i - \bar{y})^2}, \quad (6)$$

where $y_i$ is the true value, $\hat{y}_i$ is the predicted value, and $\bar{y}_i$ is equal to $\frac{1}{n} \sum_{i=1}^{n} y_i$. An $R^2$ score equal to 1 means the model is able to perfectly predict the true y values. A value of 0 means the model is completely disregarding the input features to predict random $y$ values and finally $R^2$ value may be arbitrarily worse, which results in $R^2$ values from 0 to negative infinity depending on how poor the models performance is.

**Table 2.** Metrics calculated for each number of hidden layers to determine the optimal ANN depth. For each number of hidden layers we train the ANN 10 times to find an average for MSE, MAE and $R^2$ score for which we calculate the standard error. We determine that 4 layers is the optimal number of hidden layers as this gives closest $R^2$ score to 1 and the lowest MSE and MAE.

| Layers | MSE | MAE | $R^2$ score |
| --- | --- | --- | --- |
| 1 | 0.059 ± 0.004 | 0.164 ± 0.006 | −0.07 ± 0.07 |
| 2 | 0.045 ± 0.002 | 0.159 ± 0.004 | 0.18 ± 0.04 |
| 3 | 0.040 ± 0.003 | 0.150 ± 0.005 | 0.28 ± 0.05 |
| 4 | 0.032 ± 0.001 | 0.138 ± 0.002 | 0.42 ± 0.02 |
| 5 | 0.032 ± 0.001 | 0.140 ± 0.003 | 0.41 ± 0.02 |
| 6 | 0.033 ± 0.001 | 0.140 ± 0.002 | 0.41 ± 0.02 |
| 7 | 1.2 ± 0.4 | 0.7 ± 0.2 | −21 ± 7 |

**Table 3.** Metrics calculated for the different scales of hidden nodes. We use MSE, MAE, and $R^2$ score to show that scale $a = 1$ gives the best metrics using equation (7) for which we calculate the standard error. This gives a total of 456 nodes in the hidden layers; starting with 256 hidden nodes in the first hidden layer and halving this for each subsequent hidden layer.

| Scale | MSE | MAE | $R^2$ score |
| --- | --- | --- | --- |
| 1 | 0.0326 ± 0.001 | 0.137 ± 0.003 | 0.41 ± 0.02 |
| 2 | 0.0404 ± 0.003 | 0.152 ± 0.006 | 0.26 ± 0.05 |
| 3 | 0.0504 ± 0.004 | 0.162 ± 0.008 | 0.08 ± 0.08 |
| 4 | 0.09 ± 0.03 | 0.21 ± 0.04 | −0.6 ± 0.6 |
| 5 | 0.0516 ± 0.003 | 0.164 ± 0.005 | 0.06 ± 0.06 |
| 6 | 0.14 ± 0.09 | 0.24 ± 0.08 | −2 ± 2 |
| 7 | 0.2 ± 0.2 | 0.3 ± 0.1 | −3 ± 3 |
| 8 | 0.4 ± 0.4 | 0.3 ± 0.2 | −7 ± 7 |
| 9 | 1.1 ± 0.7 | 0.6 ± 0.3 | −19 ± 13 |
| 10 | 1.8 ± 0.7 | 0.9 ± 0.3 | −31 ± 12 |

### 3.2 ANN architecture

The architecture of our ANN consists of an input layer with 14 nodes for each of the 14 input EWs, four hidden layers with a total of 465 nodes and an output layer with one node. We calculate the evaluation metrics for 1–7 hidden layers and determine 4 hidden layers perform the best as shown in Table 2. We choose 4 layers as opposed to 5 or 6 even though they have nearly the same result for all metrics because we want a network with the smallest size that achieves this best overall score as this will help to prevent over fitting.

Generally, the number of nodes in the first hidden layer will be between the input size and the output size to prevent over fitting, however, after testing various sizes of ANN we determine the best to be given by the equation

$$N_h = \frac{N_s}{a(N_i + N_o)}, \quad (7)$$

where $N_h$ is the total number of nodes in the hidden layers, $N_s$ is the number of samples in the data set, $N_i$ is the number of input nodes, $N_o$ is the number of output nodes, and $a$ is a scale factor generally found to be between 1 and 10 through testing (see supplementary material of Liu et al. 2020). We find $a = 1$ to perform the best according to our MSE, MAE and $R^2$ scores shown in Table 3. The final ANN architecture is shown in Fig. 1, the network has 14 nodes in the input layer, 4 hidden layers and 1 node in the output layer.

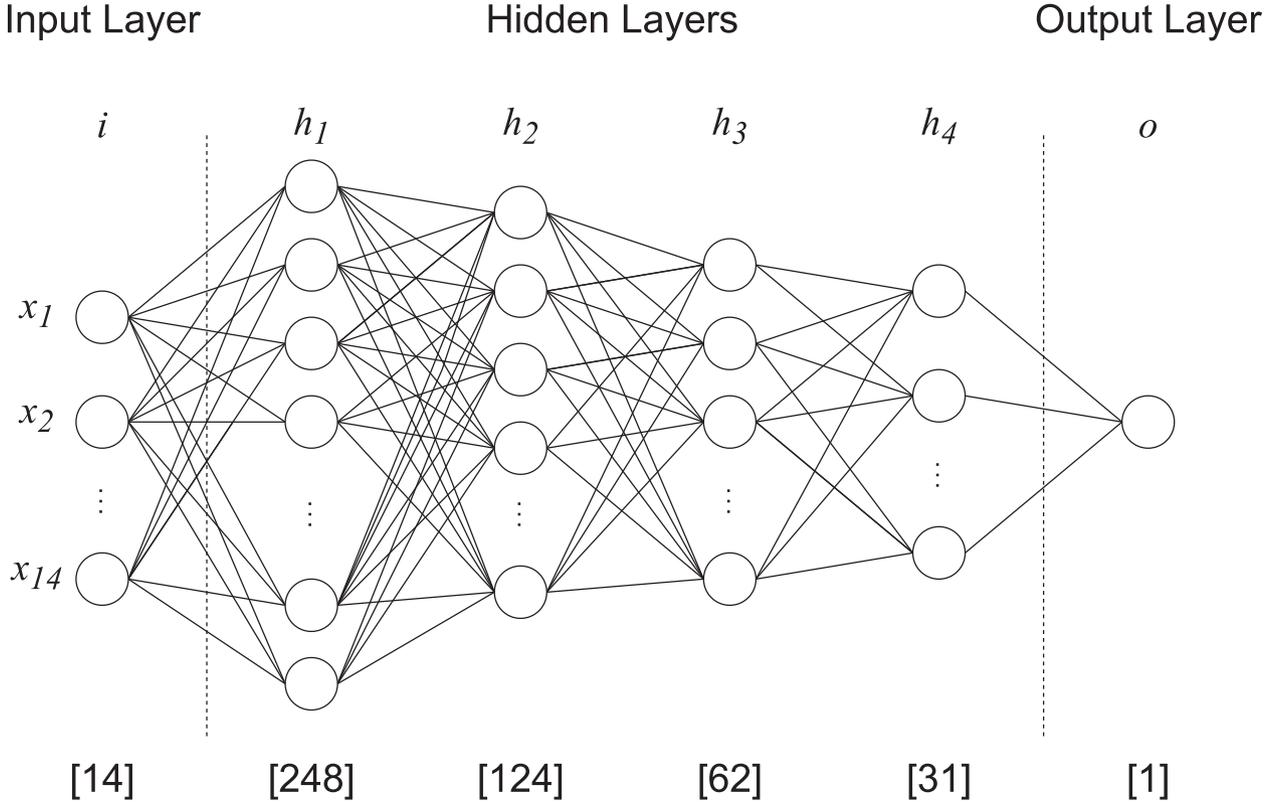

**Figure 1.** Our ANN has six layers which include an input layer ($i$), four hidden layers ($h_1$, $h_2$, $h_3$, $h_4$) and an output layer ($o$). There are 14 nodes in the input layer as we have 14 input features in the form of EWs. $h_1$ has 248 nodes and every subsequent layer has half the previous layer, such that $h_2$ has 124 nodes, $h_3$ has 62, and $h_4$ has 31 nodes. The output layer has 1 node that gives the predicted age of the galaxy based on its input EWs.

### 3.3 Hyperparameters

Here, we detail the chosen hyperparameters of our ANN. We offer descriptions of the processes used to determine which hyperparameters are the most effective for our data and model. For the activation functions in the hidden layers, the loss function and the optimizer we compare the performance of our model with various pre-built functions provided by the PYTHON package Tensorflow Keras. We compare by training the model 10 times and averaging the metrics described in subsection 3.1 similarly to how we conduct our tests for the number of hidden layers and scale of the nodes. For all of our tests we use a test-train split of 20 per cent and 80 per cent, respectively.

#### 3.3.1 Activation function

An activation function can be used in each layer of an ANN to control how the network learns the training set in a linear or non-linear capacity, enabling more complex relationships to develop (Sibi, Jones & Siddarth 2013). In simple terms, it achieves this by activating and deactivating the nodes by calculating the weighted sum and further adding bias through backpropagation. We choose two activation functions for our hidden layers and output layer to optimize the performance of our ANN. First, we choose the linear activation function provided with Tensorflow Keras for our output layer as our method requires unbounded output values for y. The linear function is calculated as $y = x$.

**Table 4.** The results of our activation function tests. Each activation function is implemented as part of the PYTHON package Tensorflow Keras and tested by training the ANN on ten separate occasions for which we average the metrics MSE, MAE, and $R^2$ score and find the standard error. Softsign performs the best but is not dissimilar to tanh.

| Activation | MSE | MAE | $R^2$ score |
|---|---|---|---|
| Linear | 0.07 ± 0.01 | 0.22 ± 0.02 | −0.3 ± 0.2 |
| ReLU | 0.10 ± 0.02 | 0.23 ± 0.03 | −0.9 ± 0.4 |
| Swish | 0.037 ± 0.003 | 0.148 ± 0.007 | 0.33 ± 0.06 |
| Sigmoid | 0.051 ± 0.002 | 0.196 ± 0.005 | 0.07 ± 0.04 |
| Softmax | 0.64 ± 0.02 | 0.76 ± 0.01 | −10.6 ± 0.3 |
| Softplus | 0.035 ± 0.001 | 0.15 ± 0.01 | 0.36 ± 0.09 |
| Softsign | 0.0252 ± 0.0001 | 0.1253 ± 0.0004 | 0.543 ± 0.002 |
| Tanh | 0.0253 ± 0.0002 | 0.1241 ± 0.0005 | 0.540 ± 0.003 |
| SeLU | 0.0301 ± 0.0009 | 0.137 ± 0.002 | 0.45 ± 0.02 |
| eLU | 0.029 ± 0.001 | 0.134 ± 0.002 | 0.47 ± 0.02 |

The activation function for the hidden layers is chosen by testing the performance of various activation functions. We train our model on 10 separate occasions for each activation function in order to find an average for the metrics described earlier. The result of these tests show that the softsign activation function performs the best, with the hyperbolic tangent (tanh) function following in a close second place as shown in Table 4.

Softsign is an s-shaped function, similar to tanh, that tends to 1 and −1. However, softsign differs from tanh as it converges polynomially rather than exponentially which reduces the impact of the vanishing gradient problem (Turian, Bergstra & Bengio 2009; Glorot & Bengio 2010; Szandała 2021). This is an issue that relates to the gradient of the loss function approaching zero when certain activation functions are used in the hidden layers. It is caused by activation functions that transform data from a large range to a smaller range such as 0 and 1 in the case of a sigmoid function. The result of the vanishing gradient problem is that the network is harder to train. However, even though the softsign function constrains data between –1 and 1, it converges polynomially which prevents the vanishing gradient problem from occurring. Softsign is implemented as part of the activation functions provided by `Tensorflow Keras`. In which they calculate softsign as

$$y = \frac{x}{1 + |x|}. \quad (8)$$

**Table 5.** The results for the MAE, MSE, mean absolute percentage error (MAPE), mean squared logarithmic error (MSLE), huber, and log cosh loss functions. Each loss function is implemented as part of the PYTHON package `Tensorflow Keras` and tested by training the ANN on ten separate occasions for which we average the metrics MSE, MAE, and $R^2$ score and find the standard error. The MSE loss function and Huber loss function have the best MSE metric at 0.025. MAPE has the best MAE at 0.122 and the MSE loss has the best $R^2$ metric at 0.539. Therefore, we choose MSE to be the loss function for our ANN.

| Loss function | MSE | MAE | $R^2$ score |
| --- | --- | --- | --- |
| MAE | $0.0260 \pm 0.0004$ | $0.125 \pm 0.001$ | $0.519 \pm 0.008$ |
| MSE | $0.0250 \pm 0.0002$ | $0.1260 \pm 0.0008$ | $0.539 \pm 0.004$ |
| MAPE | $0.0260 \pm 0.0004$ | $0.1220 \pm 0.0008$ | $0.532 \pm 0.007$ |
| MSLE | $0.09 \pm 0.03$ | $0.203 \pm 0.008$ | $-0.7 \pm 0.6$ |
| Huber | $0.0250 \pm 0.0003$ | $0.1250 \pm 0.0006$ | $0.536 \pm 0.006$ |
| Log cosh | $0.0260 \pm 0.0004$ | $0.126 \pm 0.001$ | $0.535 \pm 0.007$ |

*3.3.2 Loss function*

A loss function acts to evaluate a models performance after each training epoch. It quantifies the error between the true value and the predicted value, this then allows the optimizer to update its weights through backpropagation. The choice of loss function depends on the output data as a regression problem requires different evaluation metrics to classification problems. Again, we test each regression-appropriate loss function available with `Tensorflow Keras` and find the MSE function performs the best with our model, the results of which are shown in Table 5. The MSE loss function is calculated in the same manner as the MSE evaluation metric

$$MSE = \frac{1}{n} \sum_{i=1}^{n} (y_i - \hat{y}_i)^2. \quad (9)$$

The MSE function calculates the average of the MSEs between a sample $n$ predicted ages $\hat{y}_i$ and their corresponding true ages $y_i$ in a sample of size $n$.

The error in GAMA age estimates can be taken into account such that the ANN should penalize poor observations more than good observations. The `Tensorflow Keras` package offers capabilities to implement custom loss functions which offer a wider range of options when tuning hyperparameters. We write a simple custom loss function that builds on the MSE loss function by weighting the loss depending on the error on the age estimation from GAMA. A custom loss function can be used to apply weights to the calculated

**Table 6.** The results of our optimizer tests. Each optimizer is implemented as part of the PYTHON package `Tensorflow Keras` and tested by training the ANN on 10 separate occasions for which we average the metrics MSE, MAE, and $R^2$ score and find the standard error. Across the board the Adam optimizer has the best average metrics with an MSE of 0.0252, an MAE of 0.1241 and $R^2$ score of 0.5421.

| Optimizer | MSE | MAE | $R^2$ score |
| --- | --- | --- | --- |
| Nadam | $0.027 \pm 0.001$ | $0.128 \pm 0.003$ | $0.52 \pm 0.02$ |
| RMSprop | $0.0273 \pm 0.0006$ | $0.129 \pm 0.001$ | $0.50 \pm 0.01$ |
| Adam | $0.0252 \pm 0.0002$ | $0.1241 \pm 0.0004$ | $0.542 \pm 0.002$ |
| Adamax | $0.0257 \pm 0.0003$ | $0.1261 \pm 0.0006$ | $0.528 \pm 0.005$ |

loss such that better observations have a higher weight and therefore the loss will be reduced.

We calculate the weights based on the width of the percentiles for the mass-weighted age estimates from GAMA. We find the range between the 2.5th-97.5th percentiles and normalize between 0 and 1 such that a wider range between the percentiles corresponds to a higher weight being placed on observations with wider errors. To create the custom loss function we alter the MSE loss function by dividing the MSE by the weight for each galaxy so the loss during training will be lower for galaxies with a smaller percentile range as this indicates a lower error

$$loss = \frac{1}{n} \sum_{i=1}^{n} \frac{(y_i - \hat{y}_i)^2}{w_i}, \quad (10)$$

where $w_i$ is the weight for a galaxy with a true age $y_i$ and a predicted age $\hat{y}_i$ in a sample of size $n$. To implement this we include the weight for a given observation in the y input data by adding it as an extra dimension in the y array before applying the train test split. This is necessary given `Keras` custom loss function capabilities which only accept one input for a given observation. This minimization of loss for smaller errors acts to help the network by prioritizing the more accurate true ages and therefore, favour learning the relationships between EW and age from these observations more than poorer observations.

*3.3.3 Optimizer*

An optimizer updates the weights and learning rate of the model with the goal of reducing the loss as much as possible. We choose the stochastic gradient descent optimizer function called Adam provided with `Tensorflow Keras` (Kingma & Ba 2014). Adam stands for Adaptive Movement Estimation and is a combination of the AdaGrad optimizer's ability to deal with sparse gradients (Duchi, Hazan & Singer 2011) and RMSprop's ability to deal with non-stationary objectives (Tieleman, Hinton et al. 2012). Adam works by estimating first-order and second-order moments. We choose Adam with the same method as the other hyperparameters and detail our results in Table 6.

# 4 RESULTS

In order to fully evaluate the predictive power of the network we compare our evaluation metrics described in subsection 3.1 which allows us to directly compare the performance of the ANN with different data sets. We predict the ages for 500 randomly selected galaxies that are not included in the training or testing phases, such that the ANN has never seen the samples before. We choose a different random set of 500 galaxies 20 times in order to calculate average results for our metrics. The purpose of this is to ensure the ANN does not get lucky with a good sample of observations that causes the ANN to appear more accurate than it really is. We plot the predicted ages against the true ages from GAMA, as shown in Fig. 2, with the aim of having a perfect linear correlation in which $t_t = t_p$, where $t_t$ is the true GAMA estimated age and $t_p$ is our ANN predicted age. We find our evaluation metrics to give an average MSE of 0.020, MAE of 0.108 and $R^2$ of 0.530.

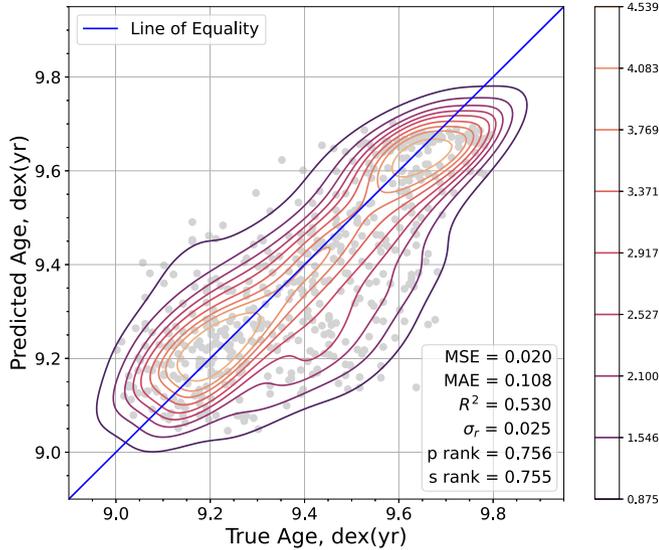

**Figure 2.** Here we show the ability of the ANN to predict galaxy ages in direct comparison with their true ages with a kernel density estimate (KDE) plot. There is some scatter in the results but a clear linear relationship. A perfect set of predictions should follow a one-to-one relationship as shown with the line of equality. The KDE contour levels represent isoproportions of density such that the most outer contour excludes 10 per cent of the probability mass and the most inner contour excludes 90 per cent. Each contour increases in 10 per cent per cent intervals between 10 and 90 per cent. The contours show that the distribution of predicted ages is skewed towards a higher true age. This shows that the ANN is underestimating older ages and overestimating younger ages.

Fig. 2 shows the ANN is able to predict the ages with appreciable accuracy. To compare the overall distribution of the predictions in comparison with the true ages, we calculate the mean and standard deviation (scatter) of each, respectively – this is with the goal of matching the distributions and means. The true ages have a mean and standard deviation of $\mu_t = 9.405$ and $\sigma_t = 0.207$, whereas the predicted ages have a mean and standard deviation equal to $\mu_p = 9.377$ and $\sigma_p = 0.182$ which gives a residual value of $\sigma_r = 0.025$, this shows that our ANN is matching the general distribution of ages but overall is underpredicating ages. The standard deviation of the true ages is larger than that of the predicted ages which indicates the network is susceptible to the phenomenon of regression toward the mean. As such, in future work it would be important to include a larger range of ages which would introduce more extreme samples to our training set. We confirm the correlation between the true and predicted ages with the Pearson and Spearman rank coefficients for which we find $p = 0.756$ and $s = 0.755$. Fig. 3 shows how the true age relates to the prediction ability of the ANN. We calculate the difference between the predicted and true ages of the galaxies and plot this against their true ages. We find that the older the true age of a galaxy is the more the ANN underestimates the predicted age which can be seen in the bimodality of Fig. 2. The affects of this underestimation appear stronger than the overestimation of the young ages. The overall affect of this is most likely because these galaxies have more extreme EWs, as shown in Fig. 7 is comparison to the intermediate aged galaxies.

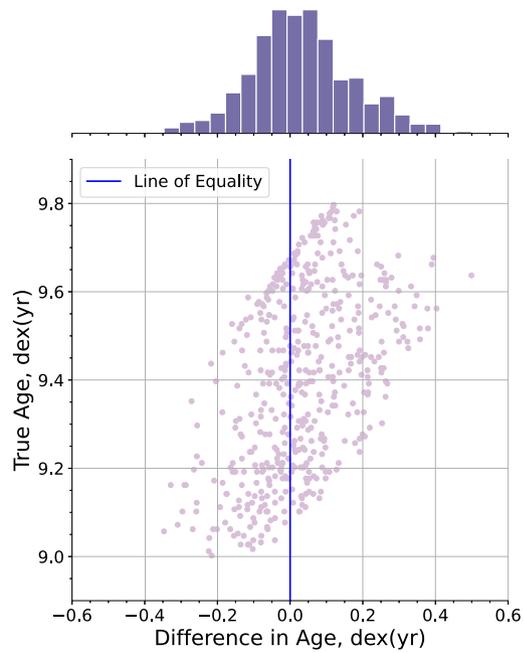

**Figure 3.** Here we represent the linked uncertainty in the true ages $t_t$ estimated by GAMA with our predicted ages $t_p$. We find the difference between the true ages and the predicted ages corresponding to Fig. 2 such that the difference in age $t_d = t_t - t_p$. The line of equality in Fig. 2 equates to a vertical line where the true ages would perfectly match the predicted ages at $x = 0$. This shows that the older galaxies are more likely to be underestimated as opposed to the younger galaxies, whereas younger galaxies are more likely to be overestimated than older galaxies. We see this in the diagonal shape apparent within the scatter plot. This appears to have an equal distribution between young and old galaxies as the histograms show an even spread of galaxies. As such, the network does not seem to overestimate young ages anymore than it underestimates older galaxies.

### 4.1 Correlation of results with properties

To determine whether the worse predictions are a result of outliers in our true data set, we plot the results of Fig. 2 coloured by different features of the data. We compare the properties sSFR, metallicity, and stellar mass in Fig. 4. High sSFR, metallicity, and stellar mass all correlate with higher ages however, the galaxies that have been more mispredicted do not seem to be extreme cases and therefore do not show any trend correlating to these properties.

Fig. 5 shows how the results correspond to the true median age percentile range estimates for each galaxy from GAMA. Similarly to Fig. 4, there appear to be extreme values of percentile width however they do not necessarily correspond to the outlying age predictions. Though the 2.5–97.5th percentile range shows a wider range for the younger galaxies which suggests the GAMA ages are less precise when their are younger stellar components. The galaxies with very narrow percentile widths tend to be close to the line of equality whereas the wider ranges tend to fall closer to the outer edges of the mispredictions, though there is a number of wide ranges towards the middle.

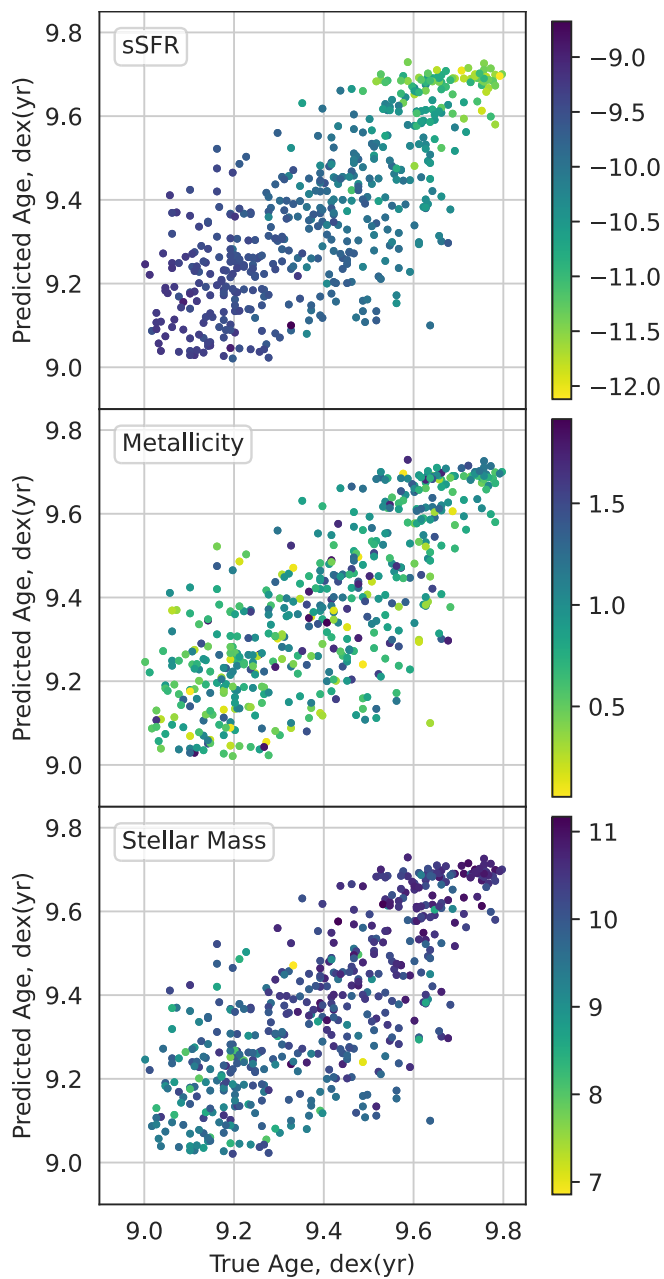

**Figure 4.** The results of Fig. 2 are coloured based on their sSFR, metallicity, and stellar mass. A darker blue corresponds to higher value for each respective property whereas light yellow shows a low value.

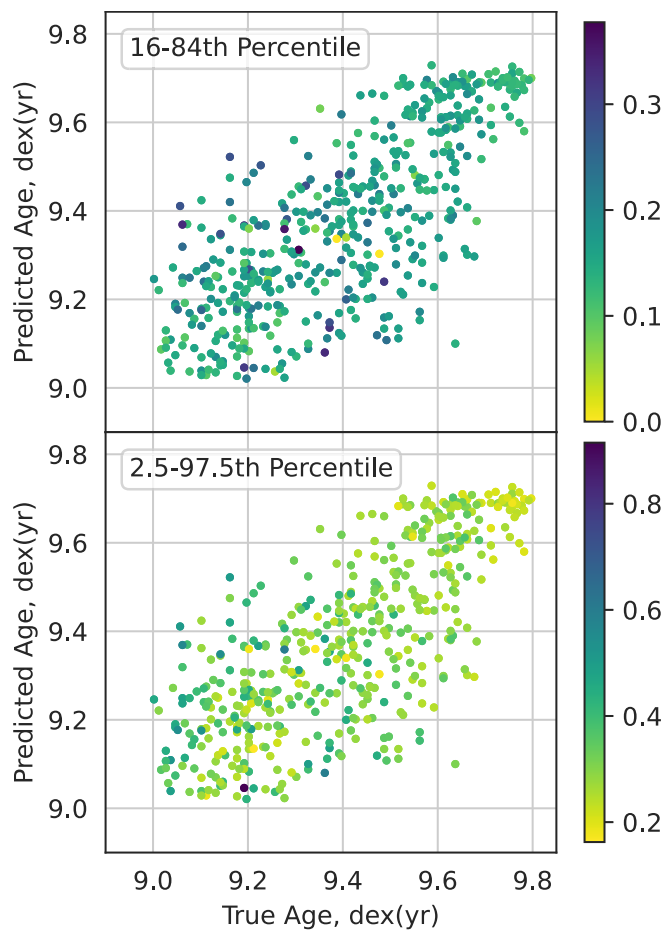

**Figure 5.** We colour the points of Fig. 2 based on the percentile range of the true mass-weighted ages from GAMA. The darker points correspond to galaxies with a wider percentile range that the true median mass-weighted age falls within. This shows that the outliers are not necessarily dependent on the precision of the true ages.

We compare the EW values for each galaxy and how they relate to predicted and true age by colouring Fig. 2 based on normalized EW values, as shown in Fig. 6. General trends can be seen across the different EWs, such as high H$\alpha$, [O III]R and [S II]B values being associated with younger galaxies. The important thing to note from this figure is that there do not appear to be outlying EW values associated with the outlying predictions. In addition, there is only one galaxy that is more than 3$\sigma$ away from the mean predicted age which is located at approximately (9.64 dex, 9.1 dex).

In order to compare how the different observations of EWs relate to the results of Fig. 2, we colour each galaxy by a its normalized SNR for each EW used in training, as show in Fig. 4. Most of the plots show that lower SNRs are associated with the worse mispredictions which suggests the network is negatively affected by poor observations of EWs. The plots for average, [O III]R, [O III]B, [S II]R, [S II]B, D4000n, and H$\alpha$ SNRs show that the high SNRs are associated with the younger galaxies whereas the low SNRs for H$\gamma_A$, G, MgG, NaD, and CNB are correlated with younger galaxies.

### 4.2 Prediction uncertainty

As discussed in Section 2, our EW input features from GAMA each have an associated observation error. As this error could introduce uncertainty into the network predictions, we calculate an aggregated proxy for uncertainty for our predicted ages by perturbing the input features within their errors before training the network and predicting the ages for the same 500 validation galaxies. We perturb the input data within its errors with a normal distribution. Once the training data is perturbed we train the network and predict the ages for 500 validation galaxies. We do this 25 times to find an average prediction uncertainty for the age for each galaxy in the validation set. This method differs from the previous method of evaluation described in Section 4 because we do not use a random 500 galaxies for every run, we use the same 500 galaxies in order to find the range in predictions and therefore find the average uncertainty across the 500 galaxies. The results of this are shown in Table 7 for each test.

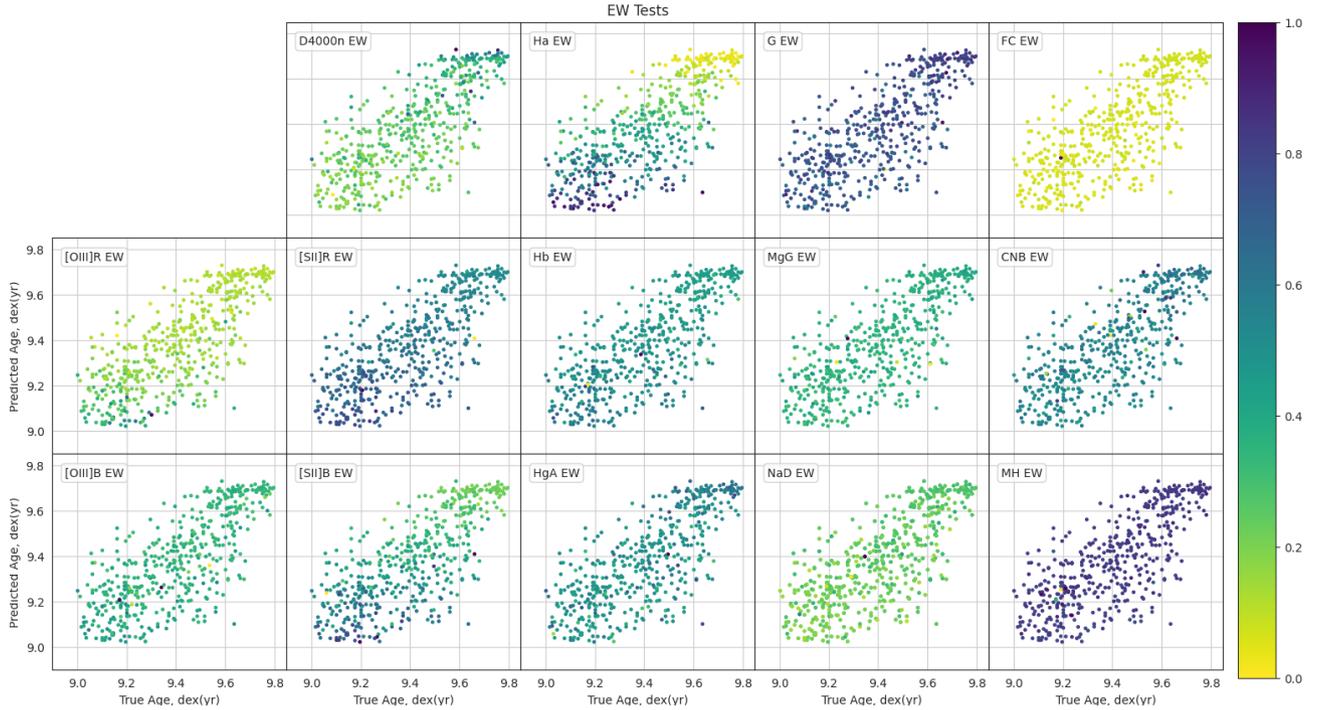

**Figure 6.** To compare the different observations of EWs with the predictions versus true ages we colour Fig. 2 based on a normalized EW for each galaxy. We normalize the EWs in order to more closely compare them.

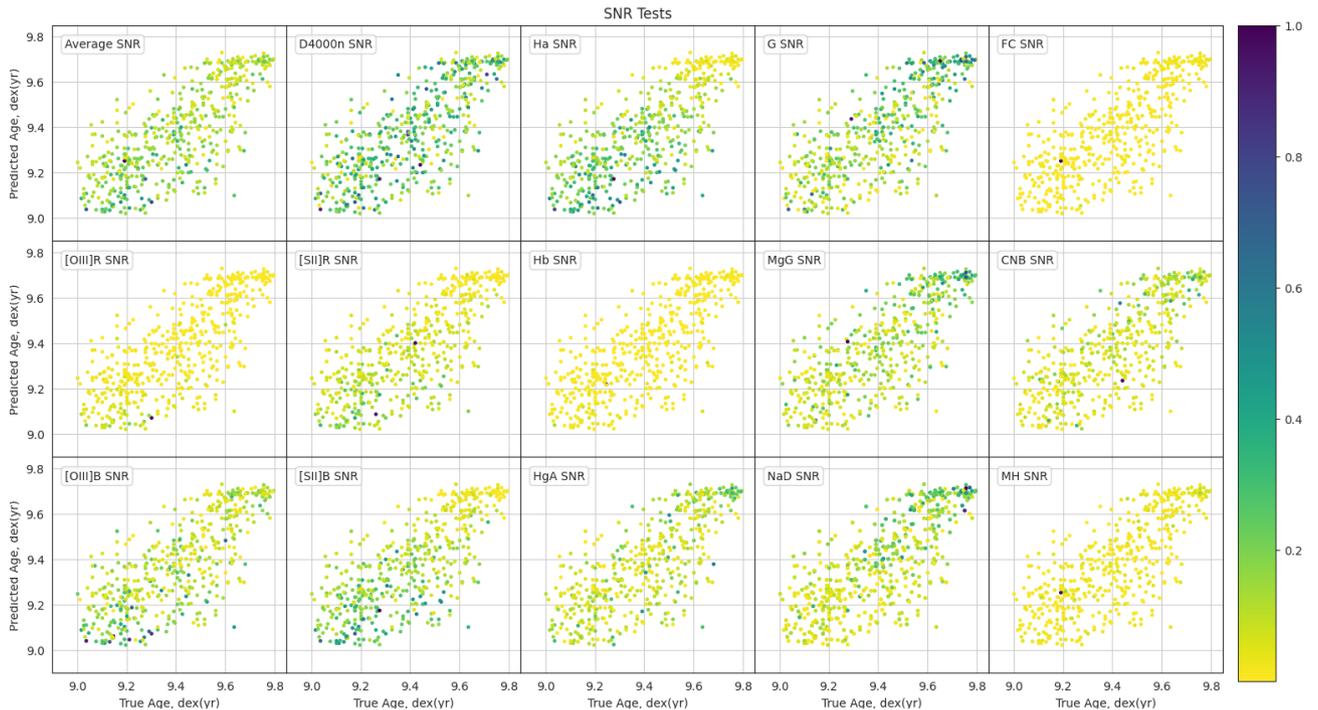

**Figure 7.** To compare the different observations of EWs with the predictions versus true ages we colour Fig. 2 based on a normalized SNR for each galaxy. We normalize the SNRs in order to more closely compare them.

The calculated uncertainty is a proxy for prediction error as this is only taking the error in the input data into account, not the error in the network predictions themselves. For this reason, we do not show the mean true age errors as this would be misleading. With this in mind the proxy for uncertainty shows that the network does not appear to be affected by fluctuations in the input data as our average prediction uncertainty and mean predicted age is $9.377 \pm 0.004$ dex (yr). This shows the network is underpredicting the overall distribution of predicted ages, though the uncertainty is quite small. We discuss this further in relation to the different evaluation tests in subsection 4.3, subsection 4.5, and subsection 5.1 in order to compare how the uncertainties are affected by different stellar populations. We also discuss how the properties of the galaxies and EW SNRs may affect this uncertainty in subsection 4.6.

**Table 7.** For both the mass- and light-weighted ages we show our prediction uncertainties. The predicted mass-weighted ages are split into the respective loss functions and sets to show how the uncertainty is affected by these different tests. The mean true age and predicted age are measured in dex (yr).

| Loss function | Set | True age | Predicted age |
|---|---|---|---|
| **Mass-weighted** | | | |
| MSE | Mixed | 9.405 | 9.377 ± 0.004 |
| MSE | Set 1 | 9.457 | 9.419 ± 0.008 |
| MSE | Set 2 | 9.355 | 9.323 ± 0.005 |
| Custom | Mixed | 9.405 | 9.368 ± 0.005 |
| Custom | Set 1 | 9.457 | 9.397 ± 0.006 |
| Custom | Set 2 | 9.355 | 9.316 ± 0.005 |
| **Light-weighted** | | | |
| MSE | Mixed | 9.380 | 9.348 ± 0.004 |

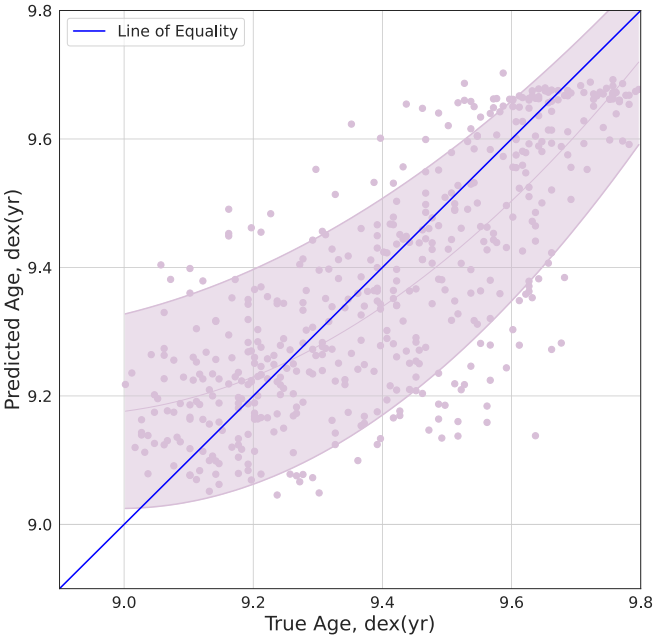

**Figure 8.** We plot the average error of the true galaxies from Fig. 2 to show how large the GAMA errors are. The mean predicted age is plotted alongside the upper and lower bounds for the average errors. This shows that though there appears to be a large amount of scatter in the predictions, they still follow the line of equality within the error bands. Therefore, the ANN is making good predictions despite the ground truth not being perfectly accurate.

### 4.3 Effect of error on age predictions

As the GAMA ages are only median estimates calculated from a percentile range there may be a large margin of error for some observations which means the ANN is being trained on ages that we do not always have accurate estimates on. To visualize the errors associated with the GAMA estimations we show the average error in Fig. 8 by plotting the mean line of best fit with upper and lower bounds to show the average error at any given age for the predicted ages corresponding to the results shown in Fig. 2. This shows that though our predictions show a large amount of scatter, they still follow the line of equality within the error band which means the ANN is making good predictions despite the error associated with the true ages.

To demonstrate this further we test the ANN by splitting the data set to calculate the results into two sets based on the percentile error of the ages. We continue to use 500 galaxies in each of the sets similarly to the method described in Section 4. We determine the sets by finding the mean range between the 2.5th and 97.5th percentiles. Any galaxies with a range less than the mean are put into Set 1, whereas galaxies with a greater range are placed in Set 2. Therefore, galaxies with a more accurate estimated age from GAMA are in Set 1 and galaxies with a greater age error are in Set 2. We ensure that all galaxies in the sets are not present in the training set.

To calculate our evaluation metrics we use a random set of 500 galaxies that the network has not seen before in order to find an average for our metrics similarly to our method in Section 4. However, to plot our figures we choose a set of 500 galaxies that produce metrics similar to that of the average. We do this to ensure the figures are comparable as the ANNs performance, in this instance, is not dependent on the selected 500 galaxies.

Fig. 11 demonstrates the performance of the network predictions by showing where predicted ages fall with respect to the true age error bars. This shows that the network is successfully predicting ages within the error of the true ages for 96.7 per cent of the total 500 galaxies, as shown in Table 11. Set 1 has a larger number of mispredictions than Set 2. We calculate 5.2 per cent of Set 1 predictions fall outside the error bars, whereas Set 2 has 1.4 per cent of predictions outside the errors. Furthermore, Set 1 has a tighter distribution within the error bars as 82.4 per cent of the predictions for Set 1 fall within the 16–84th percentile range as opposed to 78.2 per cent of Set 2 predictions, this is apparent with the more concentrated band of dark purple stars for Set 1 whereas Set 2 appears more spread out and sparse. It is interesting to note the amount of predictions outside the 16–84th percentiles but within the 2.5–97.5th is approximately the same as Set 1 has 16.6 per cent in this group and Set 2 has 16.2 per cent. This means the increased precision in Set 1 predictions that comes from the 16–84th group is being held back by the number of predictions outside the error bars and vice versa for Set 2. However, if we take into account the evaluation metrics shown in Table 8 we can see that the closer fit of the Set 1 predictions within the 16–84th error bars results in lower MSE and MAE and higher $R^2$ scores with Set 1 producing MSE = 0.018, MAE = 0.101, and $R^2$ = 0.550, whereas Set 2 has MSE = 0.023, MAE = 0.115, and $R^2$ = 0.422. The linear correlation between true and predicted age can be seen in Figs 9 and 10 for Set 1 and Set 2, respectively. Here it is apparent that there is a stronger correlation for Set 1 than Set 2 which we quantify as $p = 0.786$ and $s = 0.791$ for Set 1 and $p = 0.683$ and $s = 0.690$ for Set 2. This shows that the ANN is able to predict Set 1 ages more precisely but less reliably than Set 2 as there are more mispredictions for Set 1 than Set 2 but more accurate predictions within the 16–84th percentile range.

To compare the distribution in the predicted ages we calculate the residual scatter $\sigma_r$ which we determine to be the difference between the standard deviations of the true ages $\sigma_t$ and the predicted ages $\sigma_p$ such that $\sigma_r = \sigma_t - \sigma_p$. We find that Set 2 has a higher residual scatter with $\sigma_r = 0.032$ compared with $\sigma_r = 0.12$ for Set 1 which means the Set 2 predicted ages have a distribution closer to their true counterparts than Set 1 ages. To further investigate the distributions we find that the mean of the predicted ages of both sets is comparable to the mean of their true ages. Set 1 has a mean true age of 9.457 dex (yr) and a mean predicted age of 9.419 ± 0.008 dex (yr) whereas Set 2 has a mean true age of 9.355 dex (yr) and a mean predicted age of 9.323 ± 0.005 dex (yr). First, the mean true ages suggest that the more accurate ages from GAMA are from older galaxies which the network is able to pick up on as the mean predicted ages are similarly higher for Set 1 galaxies. The uncertainties are also comparably small for the mixed set and both Set 1 and 2, this suggests the network is not affected by fluctuations of the EWs within their uncertainties.

**Table 8.** The results for the ANN when the 500 test galaxies used to calculate the results are split into sets. Set 1 performs better as expected because the ages have a smaller error associated with them. For comparison we also include the results for a combined set. We find the mean true and predicted ages and the residual standard deviation $\sigma_r$ in order to compare the different sets as the standard deviation of the true ages varies between samples. We calculate $\sigma_r = \sigma_t - \sigma_p$ where p = predicted, t = true, and r = residual.

| Set | MSE | MAE | $R^2$ score | $\mu_t$ | $\sigma_t$ | $\mu_p$ | $\sigma_p$ | $\sigma_r$ | p rank | s rank | Time (s) |
|---|---|---|---|---|---|---|---|---|---|---|---|
| Combined | 0.020 | 0.108 | 0.530 | 9.405 | 0.207 | 9.377 | 0.182 | 0.025 | 0.756 | 0.755 | 23.52 |
| Set 1 | 0.018 | 0.101 | 0.550 | 9.457 | 0.202 | 9.419 | 0.190 | 0.012 | 0.786 | 0.791 | 25.316 |
| Set 2 | 0.023 | 0.115 | 0.422 | 9.355 | 0.198 | 9.323 | 0.166 | 0.032 | 0.683 | 0.69 | 23.624 |

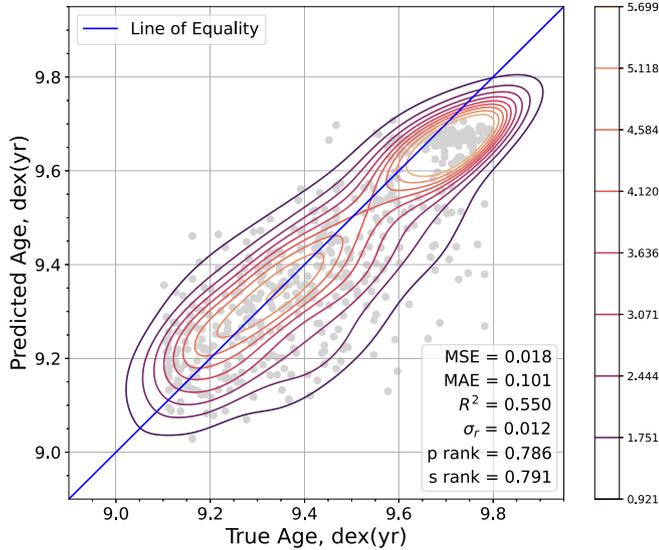

**Figure 9.** Results for predicting Set 1 ages show that the ANN is able to make more precise predictions. This is demonstrated by less scatter and narrower contours that follow the line of equality closer than those in Fig. 2 or Fig. 10. We calculate the Pearson and Spearman rank coefficients to be $p = 0.786$ and $s = 0.791$ which show a stronger correlation between the true and predicted ages for Set 1.

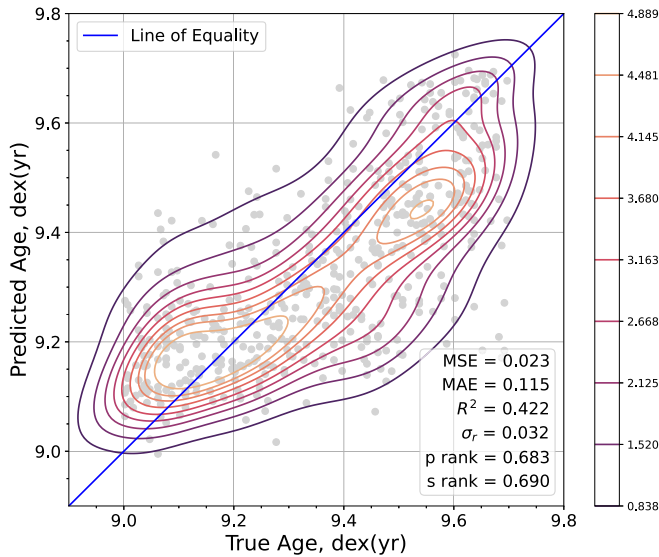

**Figure 10.** The results for Set 2 show a higher level of scatter between the true and predicted ages. This shows that the ANN performs worse when the true ages estimates have a greater error. There is still some correlation as we calculate the Pearson and Spearman rank coefficients to be $p = 0.683$ and $s = 0.690$. However, the residual scatter is greater than that of Set 1 with a value of $\sigma_r = 0.032$ for Set 2 and $\sigma_r = 0.012$ for Set 1.

One more feature in Fig. 11 to note is that the predictions tend to fall on the right side of the true age line for which we quantify in Table 12. We find that 60.2 per cent of the predictions fall on the right which means the ANN is tending towards predicting ages too low for their true age. This is most apparent for the predictions between the 2.5–16th and 84–97.5th percentiles (light purple stars) and the predictions outside the the error bars (red crosses). This can be seen in Fig. 3 as the histograms bins closest to the line of equality are fairly even but there is a higher number of points at 0.2 dex and higher. This combined with our contour plots and numerical results shows that, though the ANN is predicting more accurately in terms of numerical results and overall distribution it is struggling with some Set 1 predictions.

The ANN is therefore better at predicting the ages that have more precise estimations. To reiterate, the true ages are not necessarily exactly correct. They are estimated percentiles with a median most likely age. Thus, the ANN is able to predict the ages of galaxies more precisely for observations that have more certainty and a smaller percentile range. This is reflected in the performance of the ANN for galaxies with ages that are less certain for which the ANN predicts ages less precisely when compared with the true age estimates. This shows it is successfully finding the links between various EWs and true age as the ANN performs better for Set 1 galaxies.

### 4.4 Uncertain input data

To evaluate the performance of the network on data outside of the trained range, we test out-of-distribution ages and EWs. First, we train the network we the MSE loss function and custom loss function in order to compare the predictions. Then we use validation galaxies with ages outside the trained range, as shown in Fig. 12. The predictions for out of range ages using the MSE loss function notably perform better according to the metrics in comparison to the predictions of ages within the trained range, as shown in Table 9. The out of range predictions using the custom loss function are comparable to those within the trained age range.

Similarly, we evaluate the networks ability to predict ages for galaxies that have out-of-distribution EWs. To achieve this, we take varying portions of the validation data set and alter the EWs by increasing or decreasing the EW value by 2 times the uncertainty. We show the results of altering the EWs for 25 per cent, 50 per cent,

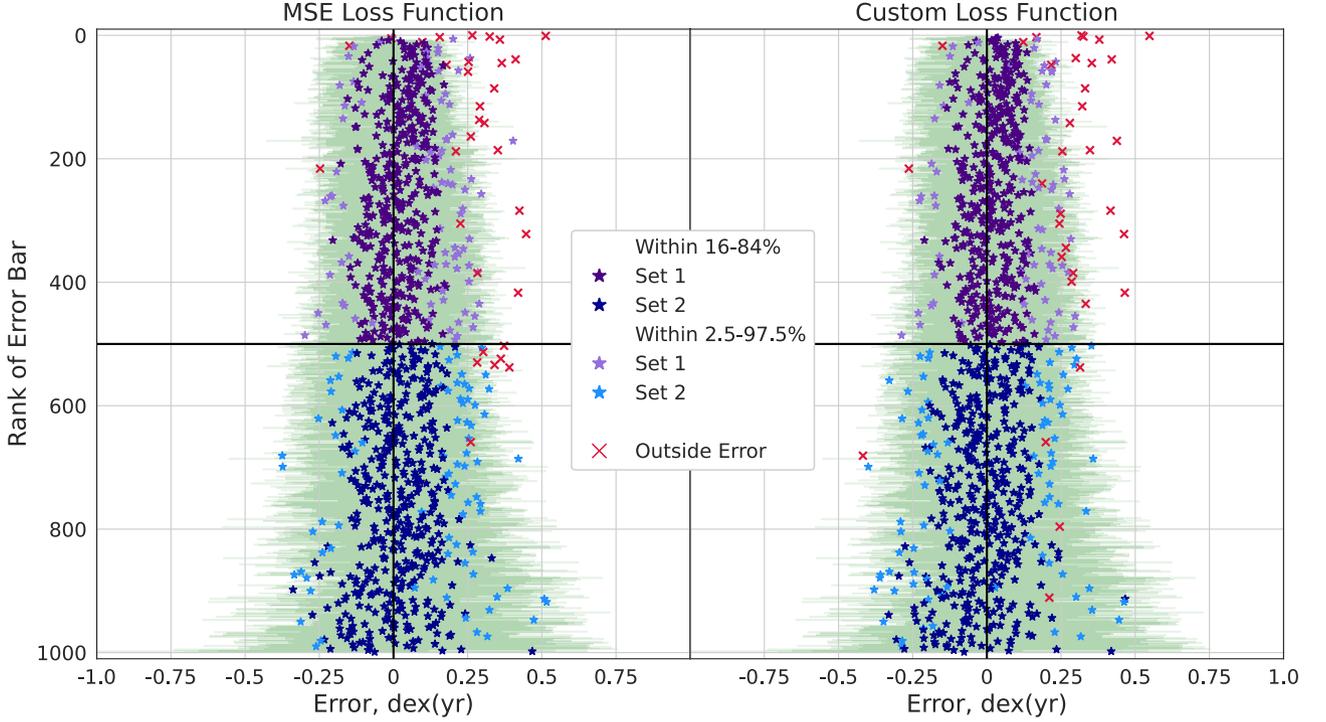

**Figure 11.** We show the predicted ages within the error bars of the true age estimates which we determine as the difference between the median mass-weighted age (true age) and the 2.5th and 97.5th percentiles as this is the maximum range for a galaxies age estimated by GAMA and as such our predicted ages should fall within. We represent the error of the true age with green error bars that are centred at 0 as to represent the true age and the *x*-axis corresponds to the difference between the true ages and the 2.7th percentile, 97.5th percentile, and predicted ages. The lower and upper limits of the error bars correspond to the difference between the median age and the 2.5th and 97.5th percentile estimates respectively, such that a galaxy with a median estimated age of 9.5 dex, 2.7th percentile equal to 9.4 dex and 97.5th percentile equal to 9.7 dex would have an error bar centred at 0, a lower error bar limit at –0.1 and an upper limit of +0.2. Predicted ages are marked with stars when they fall within the error bars and red crosses when they fall outside the error bars. Purple stars represent Set 1 galaxies as they have error bars below the average range between the 2.5–97.5th percentile range, whereas Set 2 predicted ages are shown as blue stars. The shade of purple and blue indicates whether the prediction falls within the 16–84th percentile range or the 2.5–97.5th range within the error bars. Furthermore, we rank the width of the error bars such that the narrowest range is at 1, the widest range is at 1000 and all of Set 1 is between rank 1–499 and Set 2 is between rank 500–1000. In addition, we count the number of galaxies that have predicted ages that fall outside the error bars for which Set 1 galaxies have a significant number more than Set 2. We plot the same 500 galaxies for both the MSE loss function and the custom loss function which do not show a significant difference in performance of the ANNs predictions.

75 per cent, and 100 per cent of the validation galaxies in Table 10. The evaluation metrics become considerably worse the more the validation set is altered which is apparent in the contours of Fig. 13.

### 4.5 Custom loss function

We calculate our evaluation metrics for a combined set, Set 1 and Set 2 in the same method used in subsection 4.3. Using this method we find that the ANN performs better with Set 1 than Set 2 or the combined set, as shown in Table 13. However, the overall results show that the ANN with the custom loss function is not matching the performance we see with the MSE loss function. The only metric that improves is the average time as the Set 1 average run time is 2 s faster however, both the Set 2 and combined run times are ∼2 s longer. Our evaluation metrics MSE and MAE perform similarly with and without the custom loss function with but the $R^2$ score performs worse for Set 2 and combined but slightly better for Set 1. We show the predicted ages of the combined set in Fig. 14, Set 1 in Fig. 15 and Set 2 in Fig. 16 for the same 500 galaxies used in Section 4 and subsection 4.3. The predictions for all sets appear nearly identical for the custom loss and MSE loss functions which we quantify with the Pearson and Spearman rank correlation coefficients, as shown in Table 13. The uncertainties shown in Table 7 are similarly low when compared with the uncertainties for the MSE loss function predicted ages.

The ANN is able to predict Set 1 ages more precisely than Set 2 which confirms the findings in subsection 4.3. This shows the network is able to make better predictions for galaxies that have more precise age estimates from GAMA which means the network is successfully learning the patterns between the EWs and the ages. The most likely reason that weighting the poorer observations with higher losses does not improve the predictions is that the network was already doing its best to predict patterns between EWs and ages and being told which true ages are poorer does not circumvent just how much error is associated with the data as not only the label data (true ages) has errors but also the input features (EWs).

We compare the ability of the custom loss function at increasing the accuracy of predictions within the errors from the true ages with the MSE loss function in Fig. 11. This confirms there is little difference in the predictions which indicates our ANN is already working well without the need for a custom loss function despite the error on the original data. Fig. 11 confirms that 96.6 per cent of points fall within the true age error bars which matches the predictions without the custom loss function which has 96.7 per cent of predictions within the error bars. We show the percentage distribution for the predictions

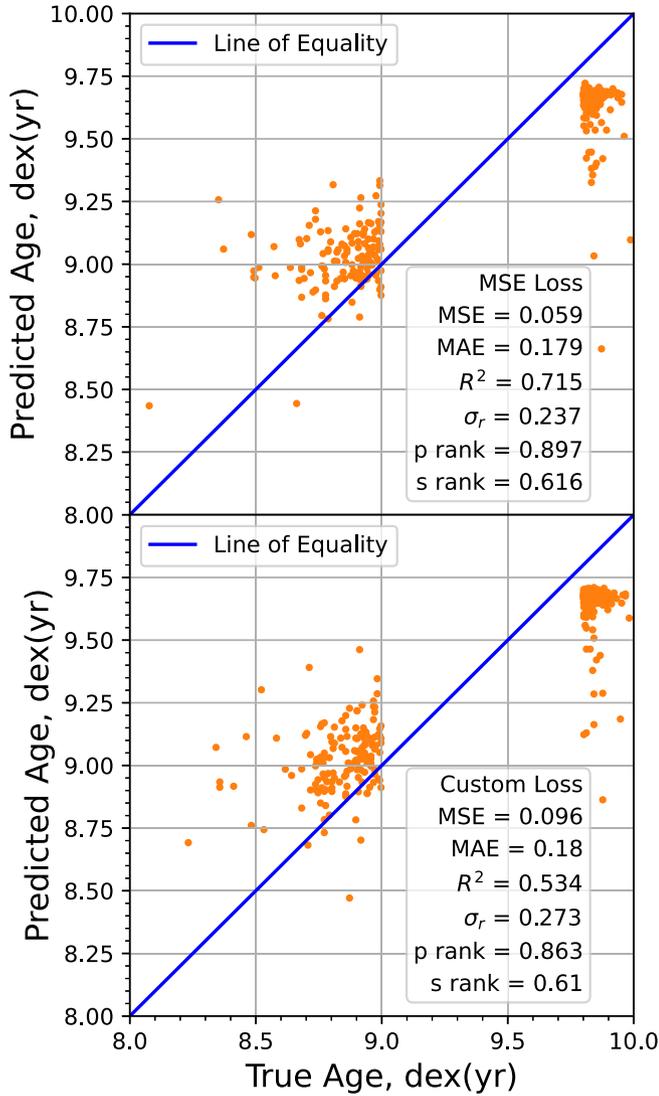

**Figure 12.** We show the predicted ages versus the corresponding true ages for galaxies outside of the trained age range. The upper panel of the figure shows the results when using the MSE loss function whereas the lower panel shows the results using a custom loss function that takes into account the error on the true ages by weighting the samples used for training.

**Table 9.** The results for the predictions on true ages that are out of the networks trained range in comparison to the results for galaxies within the trained age range. We also compare the results between the MSE loss function and our custom loss function.

| Range | MSE | MAE | $R^2$ score | $p$ rank | $s$ rank |
| --- | --- | --- | --- | --- | --- |
| MSE – in | 0.020 | 0.108 | 0.530 | 0.756 | 0.755 |
| MSE – out | 0.059 | 0.179 | 0.715 | 0.897 | 0.616 |
| Custom – in | 0.021 | 0.107 | 0.522 | 0.752 | 0.760 |
| Custom – out | 0.096 | 0.18 | 0.534 | 0.863 | 0.61 |

in comparison to the error bars for their corresponding true age in Table 11.

The main difference between the MSE and custom loss function apparent in Fig. 11 is that Set 2 has more predictions below the median true age (to the left of the 0 line) for the 16–84th percentile but similar values for outside and 2.5–97.5th such that the overall percentage is 53.4 per cent to the left, as shown in Table 12.

**Table 10.** We test the training network on a validation set that has a number of EWs outside their uncertainties. For each proportion of the validation set we calculate average values for 20 runs each.

| Proportion | MSE | MAE | $R^2$ score | $p$ rank | $s$ rank |
| --- | --- | --- | --- | --- | --- |
| 0 per cent | 0.020 | 0.108 | 0.530 | 0.756 | 0.755 |
| 25 per cent | 0.025 | 0.116 | 0.418 | 0.746 | 0.746 |
| 50 per cent | 0.026 | 0.117 | 0.391 | 0.716 | 0.742 |
| 75 per cent | 1.026 | 0.174 | −23.061 | 0.193 | 0.736 |
| 100 per cent | 1.054 | 0.177 | −23.727 | 0.191 | 0.734 |

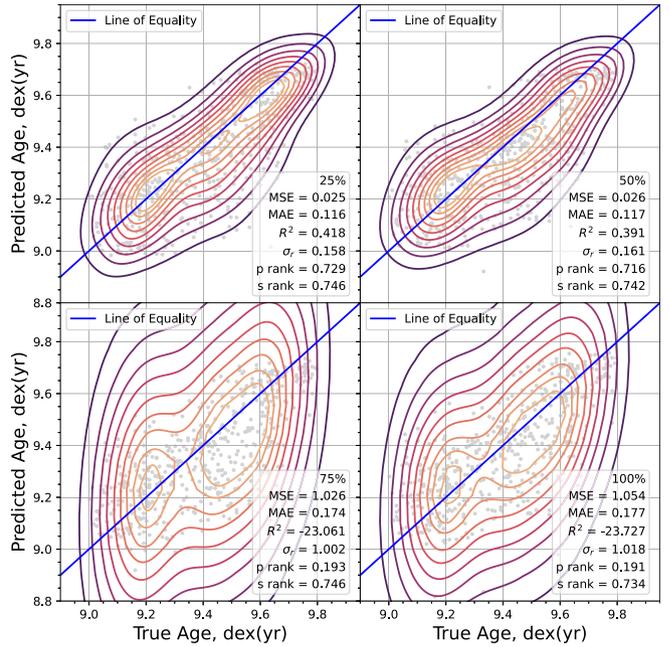

**Figure 13.** Here we show the results of altering the EWs to values outside their error ranges. The upper left and upper right panels show the results when 25 per cent and 50 per cent of the validation EWs are altered, respectively. The lower left and right panels show the results for 75 per cent and 100 per cent alteration of the validation EWs.

To summarize, our network does not seem to perform better with a custom loss function that weights the more accurate estimated ages. However, in other contexts, adding weights based on uncertainties in the data should generally improve the networks ability to learn as it is able focus on less noisy data much faster. Therefore, it is important to note that our results do not necessarily reflect those of any other network or training data. We believe that our network is already performing as well as it could with the MSE loss function such that the network is able to recognize the pattern between age and EW values. However, as the EWs also have a level of error associated with them the network could potentially be improved if this is taken into account as well. This is all to say that a custom loss function that applies weights according to the quality of the label observations may work for a different data set but as ours not only has error on the labels (estimated ages) and error on the input features (EWs) this may be preventing the custom loss function from significantly increasing the performance of the ANN.

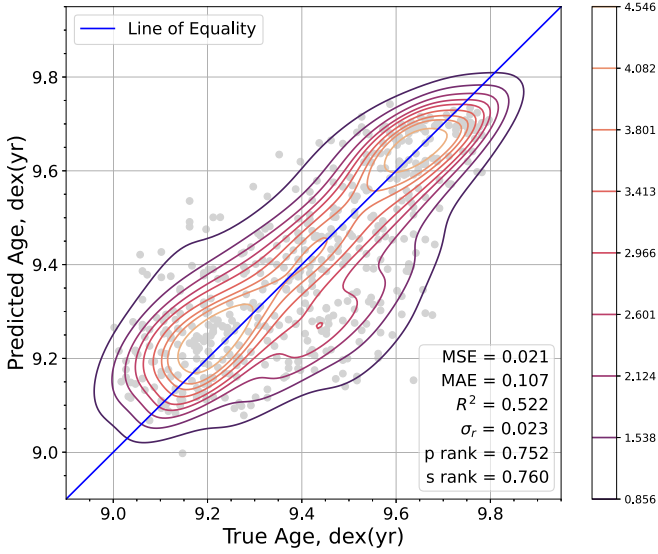

**Figure 14.** When the ANN is trained with the custom loss function there is less scatter present between the true and predicted ages. There is a strong correlation between the true ages and the predicted ages.

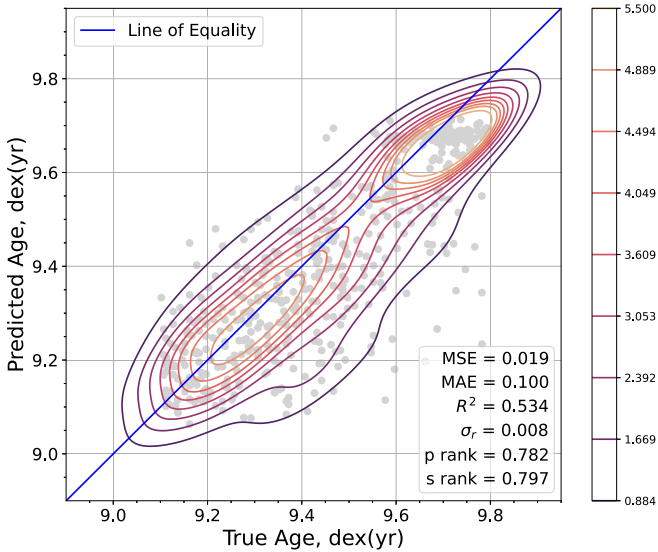

**Figure 15.** The ANN performs well when predicting the ages for galaxies present in Set 1. There is a strong positive correlation between true and predicted such that the Pearson and Spearman rank coefficients are calculated to be $p = 0.782$ and $s = 0.797$. The scatter is reduced which can be seen in the shape of the contours as they are more concentrated over the line of equality. This can be quantified with the residual scatter which we find to be $\sigma_r = 0.008$.

### 4.6 Effect of properties and SNRs on age predictions

We relate the network to physical properties of the galaxies in our data set by comparing the mean ages and uncertainties to different sets of galaxies. We use the physical properties: specific SFR and stellar mass from GAMA in the `MAGPHYS` DMU in combination with the SNRs of our input EWs. For each test we split our validation set into high and low sets based on the median value of each property. This allows us to compare our networks predictions and whether they follow the trends seen in the true ages for each set whilst comparing the prediction uncertainties for each property. This gives us insight into whether the physical properties themselves may affect the network predictions. We find that the network is able to differentiate between the high and low sets, as shown in Fig. 17. Though, the distribution of mean ages can be seen to be overall higher for Set 1 galaxies in comparison to Set 2 galaxies with an overall difference of ∼0.1 dex (yr) The prediction uncertainties are generally not affected by the difference in properties, as can be seen with the error bars for the mean predicted ages in Fig. 17.

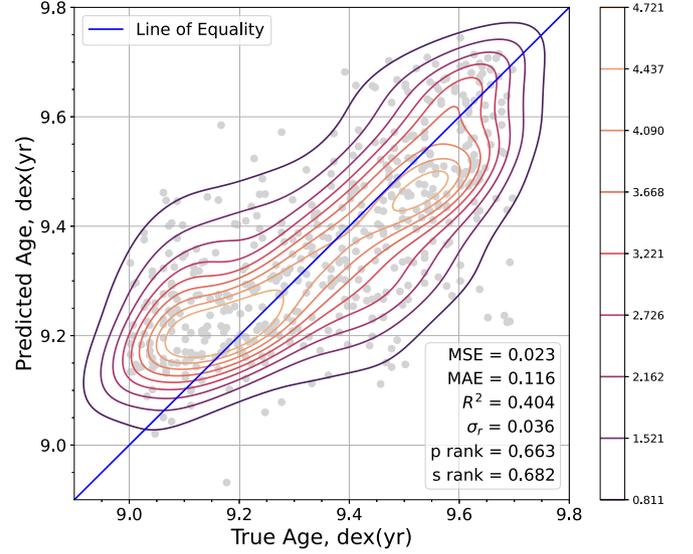

**Figure 16.** The predicted ages for Set 2 are less accurate than the Set 1 predictions. There is more scatter present between the true and predicted ages which can be quantified with the residual scatter $\sigma_r = 0.036$ which is much higher than that of Set 1 and the combined sets, as shown in Table 13.

**Table 11.** We calculate the percentage of predictions that fall within the 16–84th percentile range (In 16–84), outside the 16–84th range but within the 2.5–97.5th range (in 2.5–97.5), the total within the error bars (Total In) and total outside the error bars (Outside). We calculate this for Set 1, Set 2, and the combined sets for both the MSE and custom loss functions to compared the difference between predictions. The results correspond the predictions in Fig. 11. The percentage of predictions within each band are almost exactly the same between the MSE and custom loss functions with ∼ 96 % of predictions falling within the error bars and ∼ 80 %.

| MSE | Total in | In 16–84 | In 2.5–97.5 | Outside |
|---|---|---|---|---|
| Combined | 96.7 per cent | 80.3 per cent | 16.4 per cent | 3.3 per cent |
| Set 1 | 94.8 per cent | 78.2 per cent | 16.6 per cent | 5.2 per cent |
| Set 2 | 98.6 per cent | 82.4 per cent | 16.2 per cent | 1.4 per cent |
| **Custom** | **Total in** | **In 16–84** | **In 2.5–97.5** | **Outside** |
| Combined | 96.6 per cent | 80.4 per cent | 16.2 per cent | 3.4 per cent |
| Set 1 | 94.2 per cent | 79.2 per cent | 15 per cent | 5.8 per cent |
| Set 2 | 99 per cent | 81.6 per cent | 17.4 per cent | 1 per cent |

Higher sSFRs are associated with younger, star-forming galaxies whereas lower sSFRs are generally associated with older, quiescent galaxies in which star formation has been quenched. Fig. 17 shows this relationship clearly as there is a large difference in the mean true ages when separated by high versus low sSFR. The mean predicted ages for the high sSFR set also shows good agreement with the mean true age which suggests the network is predicting the younger ages more accurately. This is in agreement with previous discussions about the network underpredicting the older ages as the low sSFR, older set has a lower mean predicted age with a difference of 0.5 dex (yr).

**Table 12.** To demonstrate whether the network is under- or overpredicted ages we find the percentage of predictions above and below the true age similarly to the method used for Table 11. For Set 1, Set 2, and the combined set we find the percentage of predictions that fall within the error bands and the total that fall above or below their corresponding true age. If a prediction is lower than the true age it will fall to the right of the Fig. 11 as the difference will be positive, whereas predictions that are higher than their true age will fall to the left as the difference will be negative.

|             | MSE            |                | Custom         |                |
|             | Left           | Right          | Left           | Right          |
| --- | --- | --- | --- | --- |
| **Combined** | | | | |
| In 16–84    | 37.7 per cent  | 42.7 per cent  | 34.5 per cent  | 45.8 per cent  |
| In 2.5–95.7 | 6.4 per cent   | 9.8 per cent   | 5 per cent     | 11.4 per cent  |
| Total in    | 44.1 per cent  | 52.5 per cent  | 39.5 per cent  | 57.2 per cent  |
| Outside     | 0.3 per cent   | 0.8 per cent   | 0.3 per cent   | 3 per cent     |
| Total       | 44.4 per cent  | 55.6 per cent  | 39.8 per cent  | 60.2 per cent  |
| **Set 1**   | | | | |
| In 16–84    | 30.8 per cent  | 48.4 per cent  | 31.6 per cent  | 46.6 per cent  |
| In 2.5–95.7 | 4.2 per cent   | 10.8 per cent  | 4.8 per cent   | 11.8 per cent  |
| Total in    | 35 per cent    | 59.2 per cent  | 36.4 per cent  | 58.4 per cent  |
| Outside     | 0.4 per cent   | 5.4 per cent   | 0.6 per cent   | 4.6 per cent   |
| Total       | 35.4 per cent  | 64.6 per cent  | 37 per cent    | 63 per cent    |
| **Set 2**   | | | | |
| In 16–84    | 44.6 per cent  | 37 per cent    | 37.4 per cent  | 45 per cent    |
| In 2.5–95.7 | 8.6 per cent   | 8.8 per cent   | 5.2 per cent   | 11 per cent    |
| Total in    | 53.2 per cent  | 45.8 per cent  | 42.6 per cent  | 56 per cent    |
| Outside     | 0.2 per cent   | 0.8 per cent   | 0 per cent     | 1.4 per cent   |
| Total       | 53.4 per cent  | 46.6 per cent  | 42.6 per cent  | 57.4 per cent  |

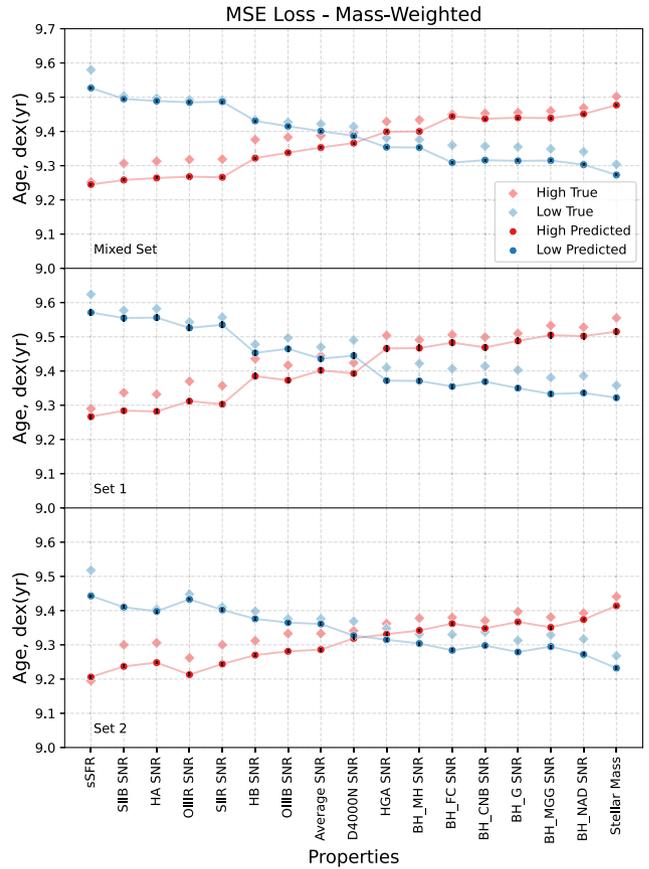

**Figure 17.** For each property listed, we split the 500 validation galaxies into a high and low sets using the median value. The order in The mean true age for the high sets are shown with pale blue diamonds whereas the mean true age for the low sets are shown with pale red diamonds. The mean predicted ages for the high and low sets are shown with blue and red circles. The prediction uncertainties are also shown with black error bars, however, these are very small in comparison to the ages. We show the mean true and predicted ages for the 17 properties for the mixed set, Set 1 and Set 2. EWs with 'BH' preceding them refer to Huchra et al. (1996) definitions.

The higher a galaxy's stellar mass, the higher their age is generally. Similarly to the sets based on high sSFR, Fig. 17 shows the distinction between high-stellar mass and low-stellar mass clearly. Again, we see the overall underprediction of all ages with a difference of 0.2 dex (yr).

Finally, we investigate how the SNR affects the network predictions by comparing the input EW SNRs and the average SNR across our 14 input features. This is to explore the networks utilization of the EWs to determine physical relationships e.g. is the network affected by the SNR for H$\alpha$ as this could affect its ability to predict younger galaxies ages as H$\alpha$ is associated with recent star formation. As shown in Fig. 17, higher SNRs for H$\alpha$, [S II]B, [S II]R, and [O III]R are significantly associated with the younger galaxies in our validation set. Less significantly, the higher average SNR, D4000$_n$, [O III]B, and H$\beta$ SNRs are also associated with younger galaxies. Whereas higher SNRs for MgG, G, and NaD are significantly related to the older galaxies, and less significantly H$\gamma_A$, MH, FC, and CNB.

Additionally, the sets associated with lower SNRs match the mean true and predicted ages more frequently whereas the higher SNRs more commonly have discrepancy between the true and predicted mean ages and overall are generally more underpredicted. When taking into account the sets described in subsection 4.3 we can see that overall the Set 1 galaxies are generally older than the Set 2 galaxies.

The prediction uncertainties described in subsection 4.2 are not affected by particular properties or SNRs as they remain unchanged across the different sets, as shown with the error bars in Fig. 17. This shows that fluctuating the EWs within their errors does not have an affect on the predictions when separated by set. In terms of the difference in between the mean true and predicted ages for each set we can see that for most of the property sets, the set that is overall older in age tends to have more accurate predictions as the mean values show less discrepancy.

**Table 13.** Results for the ANN predictions when incorporating the custom loss function. Set 1 performs better than Set 2 and the combined sets in all evaluation metrics; however, the combined set performs the best with very similar results predicted without the custom loss function. We find mean ages and the residual standard deviation $\sigma_r$ in order to compare the different sets as the standard deviation of the true ages varies between samples. We calculate $\sigma_r = \sigma_t - \sigma_p$, where p = predicted, t = true and r = residual.

| Set | MSE | MAE | $R^2$ score | $\mu_t$ | $\sigma_t$ | $\mu_p$ | $\sigma_p$ | $\sigma_r$ | $p$ rank | $s$ rank | Time (s) |
| --- | --- | --- | --- | --- | --- | --- | --- | --- | --- | --- | --- |
| Combined | 0.021 | 0.107 | 0.522 | 9.405 | 0.208 | 9.368 | 0.185 | 0.023 | 0.752 | 0.760 | 25.594 |
| Set 1 | 0.019 | 0.100 | 0.534 | 9.457 | 0.203 | 9.397 | 0.195 | 0.008 | 0.782 | 0.797 | 22.908 |
| Set 2 | 0.023 | 0.116 | 0.404 | 9.355 | 0.196 | 9.316 | 0.160 | 0.036 | 0.663 | 0.682 | 25.005 |

**Table 14.** Comparison between the performance of the ANN when trained on mass-weighted ages versus light-weighted ages. We compare the MSE, MAE, $R^2$ score, the scatter of the predicted in comparison with the true ages, the Pearson rank coefficient $p$, the Spearman rank coefficient $s$ and the total time taken by the ANN to train and predict 500 galaxies. In all regards the ANN performs better when trained with light-weighted ages.

| Age | MSE | MAE | $R^2$ score | $\mu_t$ | $\sigma_t$ | $\mu_p$ | $\sigma_p$ | $\sigma_r$ | $p$ rank | $s$ rank | Time (s) |
|---|---|---|---|---|---|---|---|---|---|---|---|
| Mass weighted | 0.020 | 0.108 | 0.530 | 9.405 | 0.207 | 9.377 | 0.182 | 0.025 | 0.756 | 0.755 | 23.52 |
| Light weighted | 0.015 | 0.094 | 0.643 | 9.380 | 0.204 | 9.348 | 0.168 | 0.036 | 0.808 | 0.836 | 24.25 |

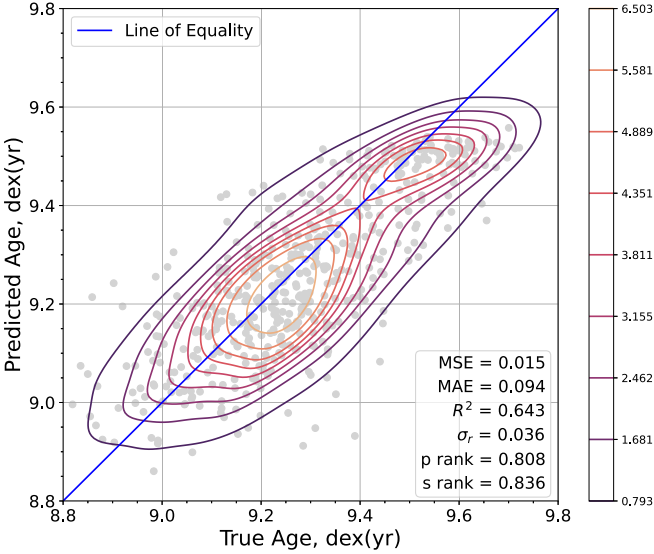

**Figure 18.** The ANN predicts light-weighted ages more accurately than mass-weighted ages. This can be seen in the overall relationship as the shape of the contours more closely follows the line of equality as opposed to the mass-weighted ages. We quantify this with the Pearson and Spearman rank coefficients which we find to be $p = 0.808$ and $s = 0.836$, respectively. The contours show that the ANN is no longer underestimating old ages and overestimating young ages like it does when it is trained on mass-weighted ages.

In conclusion, the network is able to replicate the trends we see in the properties and SNRs of young and old galaxies (e.g. high sSFR, H$\alpha$, [S II]B, [S II]R, and [O III]R for young galaxies and high-stellar mass and SNRs of MgG, G, and NaD).

## 5 DISCUSSION

We have demonstrated that our ANN is successful at predicting the ages of galaxies based on their EWs despite the true ages having a significant errors associated with them. To thoroughly evaluate our predictions we compare with the results of GAMA and other ML techniques trained on emission lines or used for stellar age prediction.

### 5.1 Mass- versus light-weighted ages

First, we discuss the affect of using mass- or light-weighted ages as discussed in Section 4. The ANN is underpredicting the ages of the older galaxies and overpredicting the younger ones, for which the results are shown in Fig. 3. This may be due to the difference in how mass-weighted ages are calculated in comparison to light-weighted ages. Mass-weighted ages are dependent on the stellar mass whereas light-weighted ages are dependent on the flux at a given wavelength. As such, light-weighted ages are biased by younger populations that dominate the luminosity but contribute very little to the total mass (Trager et al. 2000; Conroy 2013; Citro et al. 2016).

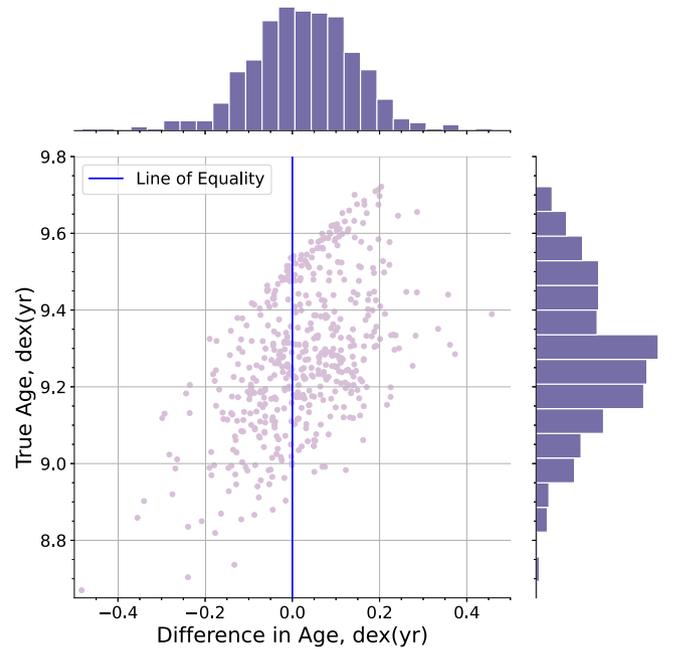

**Figure 19.** The difference in predicted light-weighted age when compared with true age shows a more even distribution of predictions. The ANN does not show a preference for underestimating old ages and overestimating young ages like it does when it is trained on mass-weighted ages.

For this reason, we investigate whether the network picks up on the differences between light- and mass-weighted ages. However, due to light-weighted ages being more sensitive to recent star formation rather than the full SFH they are easier for the network to train with but less physically meaningful. To test this we use the restframe $i$-band luminosity-weighted mean stellar ages provide by GAMA in the StellarMasses v19 DMU. The light-weighted ages perform better in all evaluation metrics, as shown in Table 14. We calculate our evaluation metrics of the light-weighted age predictions to be MSE = 0.015, MAE = 0.094, and $R^2$ score = 0.643.

To show the performance of the ANN we plot the contour graph of the true ages against the predicted ages according to the method described in Section 4. We predict the ages of 500 random galaxies that the ANN has not been trained or tested on, as shown in Fig. 18. The light-weighted ages are being predicted with less bias towards underpredicting than the mass-weighted ages as there is a more even distribution which is apparent in the shape of the contours. We demonstrate this further in Fig. 19, in which we plot the difference between the true and predicted ages against the true ages. When compared with the mass-weighted ages it can be seen that the ANN is no longer underpredicting the older galaxies and overpredicting the younger galaxies when trained on the light-weighted ages. This supports our theory that this skew is caused by the biases related to light-weighted ages with recent star formation are skewed towards younger values. To compare the scatter between the true and predicted ages we calculate the mean predicted age and standard deviation as $\mu_t = 9.380$ $\sigma_t = 0.204$ and $\mu_p = 9.348$ and $\sigma_p = 0.168$. We find the light-weighted ages have a residual standard deviation of $\sigma_r = 0.036$ which shows a smaller distribution in the predicted ages than the predicted mass-weighted ages though the mean of the light-weighted predicted ages more closely matches that of the true ages. Therefore, the network is not overall underpredicating the light-weighted ages as much as the mass-weighted ages however it is still narrowing the prediction distribution. Finally, we compare the Pearson and Spearman rank coefficients which we find to be $p = 0.808$ and $s = 0.836$, respectively.

### 5.2 Comparison to other ML algorithms

Here, we compare our ANN to other ML algorithms that have similar purposes to ours. This is due to a lack of literature describing ANNs that specifically predict the ages of galaxies. Therefore, we compare our ANN with predictions of galaxy ages via other ML techniques.

The AdaBoost and Decision Tree based ML algorithm described by Ucci et al. (2017, 2018) successfully predicts the physical properties of galaxies (density, metallicity, column density, and ionization parameter) based on their emission lines. They note that GAME still performs well when 80 per cent of their emission lines are discarded due to weak observations. This supports our decision to remove 10 of the 14 good EWs in favour of increased prediction performance and is in agreement with the increased training performance when sample weights are placed on the EWs with higher SNRs. They achieve good training times with their algorithm such that processing the SDSS DR5 would take approximately 417 h.

To compare with the CNN described by Li et al. (2022), they find a scatter of ±20 per cent between their true and predicted values. Specifically, they find stellar age and SFR reconstructed with a population wide scatter of 20 − 50 per cent. However, their average stellar masses are predicted to be 0.09 dex more accurate. Again, it is interesting to note that they find all predicted properties are slightly underpredicted.

For comparison with the CNN described by (Liew-Cain et al. 2021) it is important to note that they use two different sets of data for their training phase, Set A and Set B. Their Set A is intended to mimic a large, diverse number of galaxies such that all galactic evolutionary history is covered. Set B is more realistic as it is made from a random selection of galaxies which acts as a true survey would, as there is no previous knowledge of the data set. Their results show a standard deviation between the true and predicted ages and metallicities are $\sigma = 0.03$ for Set A which is considerably better than Set B for which they achieve a standard deviation of $\sigma = 0.16$ for both age and metallicity. They use the Pearson's correlation coefficient to evaluate the age and metallicity residuals for which they find a value of $p = -0.24$ for both Set A and Set B. This demonstrates that the model is able to make predictions that are no more affected by the age–metallicity degeneracy than the true values found with full spectral fitting. It is worth noting that their true versus predicted figures also show significant scatter in the results. It is interesting to note that not only are their estimates for Set B more spread out and scattered but they state their CNN is systematically predicting lower ages than the true spectroscopic ages. They go on to suggest that the reason for this discrepancy may lie with the fact that there is less diversity in the SFHs in Set B which means the CNN is unable to derive enough patterns to be able to predict unseen galaxies of different types. They suggest this may be resolved in future works with a larger data set or synthetically increasing the diversity of the training set.

## 6 CONCLUSIONS

We present a successful proof of concept for an ANN that is able to predict the ages of galaxies based off their spectral EWs. Our key findings are detailed below

(i) we calculated a proxy for prediction error by perturbing our input data which gives an uncertainty of ±0.004 dex (yr).

(ii) We confirm a strong positive correlation between the true and predicted ages by quantifying this relationship with the Pearson rank coefficient $p = 0.756$ and the Spearman rank coefficient $s = 0.755$. Our ANN achieves these results with a total training and predicting time of ∼23 s for 500 galaxies.

(iii) We show that the ANN is able to predict the ages of more accurate estimates better than less certain estimates by splitting our data into two sets based on the percentile range the GAMA age estimates have. This means the ANN is able to pick up on patterns between the EWs in order to predict ages which shows it is predicting based on quantifiable differences between observations rather than randomly.

(iv) We determine that weighting the loss for observations based on how accurate the age percentiles are does not significantly improve the performance of the ANN. This is not to say that weighted loss functions do not work for all data, but for our specific data set the errors on even the most accurate ages are too broad for the weighting to make a difference to the ANN. To see better results from the custom loss function we would hope that future work would focus on the quality of data rather than the amount of it.

(v) We relate our network to the physical proprieties of the galaxies and the effects of SNRs.

(vi) When trained with light-weighted ages the ANNs accuracy improves which we demonstrate by calculating the residual standard deviation between the true ages and the predicted ages. When trained with mass-weighted ages the ANN has a residual standard deviation of $\sigma_r = 0.025$ between the predictions and their true ages however, when we train with light-weighted ages the residual decreases to $\sigma_r = 0.036$. This shows that the ANN is more restricted in its distribution of light-weighted ages.

(vii) The ANN is able to predict the ages of galaxies much faster than traditional models whilst retaining comparable accuracy. This could be invaluable for future studies that only require a handful of properties such as age but a large sample of objects. A simple neural network like an ANN could be easily implemented to produce these kinds of data sets for future studies.

(viii) We compare the predictive performance of our ANN with other ML algorithms. We consider ANNs to be powerful predictors as they are fairly simple to code in Python with the ML packages `Tensorflow` and `Keras` which are able to be run on most computers.

To conclude, our method of predicting ages is a promising technique for future studies as an alternative to full simulation modelling. However, now that a proof of concept has been established, it is important to account for systematic biases of the underlying SED fitting used to produce the ages and from the network itself to account for the underprediction of older galaxies and overprediction of younger galaxies and overall underprediction of all ages. The studies that would benefit most from ANNs would be smaller studies that only require one or two properties for large sets of data for which modelling would take too long. These uses may also include determining the ages for large data sets such as SDSS in order to further study of specific age categories. This would aid in the determination of galactic evolution and formation or other studies involving ML algorithms that require an even split of ages for the prediction of other physical properties of galaxies.

# 7 DATA AVAILABILITY

The data underlying this article will be shared on reasonable request to the corresponding author.

This paper has been typeset from a TeX/LaTeX file prepared by the author.